\documentstyle[12pt]{article} 
\begin{document} 
\thispagestyle{empty} 
\begin{flushright}
UA/NPPS-1-1998
\end{flushright}
\begin{center}
{\large{\bf AN EXTENSION OF THE STATISTICAL BOOTSTRAP\\
MODEL TO INCLUDE STRANGENESS.\\ 
IMPLICATIONS ON PARTICLE RATIOS\\}} 
\vspace{2cm} 
{\large A. S. Kapoyannis, C. N. Ktorides and A. D. Panagiotou}\\ 
\smallskip 
{\it University of Athens, Division of Nuclear and Particle Physics,\\ 
GR-15771 Athina, Hellas}\\ 
\vspace{1.0cm}
{\it Accepted for Publication in Physical Review D}
\vspace{1.0cm}
\end{center}
 
\begin{abstract} 
The Statistical Bootstrap Model (SBM) is extended to describe hadronic 
systems which carry the quantum number of strangeness. The study is 
conducted in the three-dimensional space $(T,\mu_q,\mu_s)$ of 
temperature, up-down and strange chemical potentials, respectively, wherein 
the existence of a ``critical'' surface is established, which sets the limits of the 
hadronic phase of matter. A second surface, defined by the null expectation 
value of strangeness number ($<S>=0$) is also determined. The approach of 
the latter surface to the critical one becomes the focal point of the present 
considerations. Two different versions of the extended SBM are examined, 
corresponding to the values 2 and 4 for the exponent $\alpha$, which 
determines the asymptotic fall-off of the mass spectrum $\rho(m)$. It is 
found that the $\alpha=4$ version has decisive physical advantages. This 
model is subsequently adopted to discuss (strange) particle ratios pertaining 
to multiparticle production processes, for which a thermal equilibrium mode 
of description applies. 
\end{abstract} 
\vspace{1cm}

\newpage
\setcounter{page}{1} 
{\large{\bf 1. Introduction}} 
 
The Statistical Bootstrap Model (SBM) [1,2,3] constitutes an effort for a self 
consistent thermodynamical description of relativistic, multiparticle systems. 
The basic idea is to bypass the employment of interaction between particles 
at a distance, in favour of successive levels of organisation of matter into 
particle-like entities of increasing complexity known as {\bf fireballs}. In the 
context of strong interaction physics, the original set of {\bf input} particles 
corresponds to all known hadrons. 
 
The quantity which carries dynamical information in the bootstrap scheme is 
the mass spectrum $\rho(m)$ of the fireballs. It satisfies an integral equation 
with the generic form [4]: 
\begin{equation} 
\rho(m)=\delta(m-m_0)+\sum_{n=2}^{\infty}\int\frac{1}{n!} 
\delta\left(m-\sum_{i=1}^n m_i\right)\prod_{i=1}^n\rho(m_i)dm_i\;, 
\end{equation} 
where $m_0$ represents the mass of an input particle and $m_i$, 
$i=1,2,\ldots,\infty$, stands for the fireball masses in ascending order of 
complexity. 

The mass spectrum determines the number of particle/fireball states, with
rest- frame volume $V$, which reside inside the (infinitesimal) momentum 
region $d^3p$ around $\vec{p}$, via the quantity 
$\frac{\textstyle V}{\textstyle h^3}d^3p\cdot\int\rho(m)dm$.
A relativistic casting of this
expression calls for the introduction of Touchek's integration measure [5], 
as well as the adoption of a mass spectrum function $\tau(m^2)$ which 
depends on the relativistic invariant variable $m^2$. In the system of units 
where $\hbar=c=k=1$, to be adopted through this work, we write 
\begin{equation} 
\frac{V}{h^3}d^3p\cdot\int\rho(m)dm\rightarrow 
\frac{2V^\mu p_\mu}{2\pi^3}\int\delta_0(p^2-m^2)d^4p\cdot 
\tau(m^2)dm^2\;, 
\end{equation} 
with $\tau(m^2)$ defined by 
\begin{equation} 
\rho(m)dm=\tau(m^2)dm^2\;. 
\end{equation} 
 
The four-volume $V^\mu$ is taken parallel to the four momentum of the 
particle/fireball, i.e. 
\begin{equation} 
V^\mu=\frac{V}{m}p^\mu\;. 
\end{equation} 
Such a choice leads to the introduction of the quantity 
\begin{equation} 
B(p^2)=\frac{2V^\mu p_\mu}{(2\pi)^3}=\frac{2Vm}{(2\pi)^3}\;, 
\end{equation} 
which will play an important role in this work, with respect to the 
phenomenological input entering our analysis. 
 
The above basic features of the bootstrap scheme, when combined with a set 
of relations which account for statistical aspects of the system, lead to a self- 
consistent theoretical framework for describing the hadronic world and 
define the SBM. It was in the context of this approach that the idea of the 
existence of a critical temperature, known as Hagedorn temperature, beyond 
which the hadronic phase of matter ceases to maintain this self-consistent 
description, equivalently, ceases to exist, was first introduced [1]. 

An important accompanying issue raised by the above result was whether a
{\it new} phase of matter can be realised beyond the Hagedorn temperature. 
Criteria have been established, within the SBM [6], which address 
themselves to this question and which will be explicitly referred to in the 
main text. Nowadays, of course, with QCD having emerged as the 
universally accepted microscopic theory for the strong interaction, the 
verdict on this matter has become clear. Lattice QCD computations [7] show 
that a quark gluon plasma (QGP) phase of matter should emerge under 
extreme conditions of high temperature and/or high density. 
 
The intensive activity, both on the phenomenological and the experimental 
front, which was instigated by these non-perturbative QCD results has led to 
searches for the identification of specific signatures associated with the QGP 
state of matter. The central significance of the strangeness quantum number 
was soon recognised in connection with both the expected formation and 
subsequent hadronisation of the QGP in the ultrarelativistic nucleus-nucleus 
collisions. 
 
Most theoretical attempts to study such issues approach the QGP system 
from the microscopic side, i.e. from a viewpoint wherein quark and gluon 
degrees of freedom are adopted as basic input. In the present paper we 
undertake a corresponding investigation from the hadronic-matter side via 
the employment of the SBM. We consider such an approach worth pursuing 
both because it should complement the theoretical attempts which originate 
from the microscopic side and, more importantly, because it simulates the 
actual experimental setting in nucleus-nucleus collisions, wherein both initial 
and final stages belong to the hadronic domain. 
 
Past studies in the framework of the SBM have almost exclusively contained 
themselves to the non-strange hadronic sector. A first discussion of the 
ramifications brought onto the mass spectrum equation, cf. (1), by the 
inclusion of additional quantum numbers (strangeness included) has been 
conducted several years ago in Ref. [8]. In the present work we undertake as our 
first task to systematically extend the SBM so as to create a fully 
thermodynamical framework which includes strangeness. This we 
accomplish in Section 2, where we also discuss the connection of the 
bootstrap scheme with QCD phenomenology, in the context of the MIT bag 
model [9]. We identify two cases of specific interest which are labelled by 
corresponding values of an exponent ``$\alpha$'' that enters the asymptotic 
behaviour of the mass spectrum $\rho(m)$, as $m\rightarrow\infty$. The 
case, which has been extensively studied in the past, corresponds to 
$\alpha=2$. In Section 3 we confront the analogous investigation for our 
extended version of the SBM, while in Section 4 we do the same for the case 
$\alpha=4$. In Section 5 we present the results of our numerical study of the 
two versions of the strangeness incorporating SBM, which strongly advocate 
the case for $\alpha=4$. On the basis of this choice we take up, in Section 6, 
the issue of particle production ratios for species which carry the strangeness 
quantum number. Taking into account the zero strangeness constraint, we 
present the results of numerical analyses giving particle ratios as a function 
of temperature $T$ along a fixed direction in the $(T,\mu_q)$ plane, where 
$\mu_q$ stands for up-down chemical potential. We also provide ``critical 
values'' for these ratios, i.e. values on the intersection of the zero strangeness 
surface with the critical one, as functions of the critical temperature. Finally, 
the way to confront experimental data on particle ratios, with respect to how 
close they come to and/or whether they may have already reached the QGP 
phase, is briefly discussed. A summary of our work in this paper is presented 
in Section 7. 
 
An abbreviated account of a part of the present work has been presented in a 
recent symposium on Strangeness in Quark Matter [10]. Here we exhibit the 
computational details behind our (aforementioned) contribution, as well as 
furnish new results pertaining to numerical procedure and particle 
production ratios.

\vspace{2cm} 
{\large{\bf 2. Extension of the SBM to include strangeness}} 
 
In this Section we construct an extention of the SBM which incorporates 
strangeness as an independent quantum number. We shall present the basic 
set of equations which comprise the model, both on the bootstrap and the 
statistical front. 
 
Let us begin by fixing our notational conventions with respect to particle 
labelling on fugacity (or chemical potential) variables. We reserve the index 
``$\rm a$'' for individual hadrons. Thus, $\lambda_{\rm a}$ ($\lambda_{\rm a}^{-1}$) 
represents the fugacity of a given particle (antiparticle) belonging to the 
hadronic matter. Baryon number is denoted by ``$b$'', whereas the 
corresponding fugacity index is ``$B$'', i.e. we write $\lambda_B$. A similar 
convention is used for the, thermodynamically dual, variables pertaining to 
strangeness. Specifically, ``$S$'' stands for strangeness number and 
$\lambda_S$ is the corresponding fugacity. Finally, by $\lambda_q$ we 
denote up and down quark fugacity and by $\lambda_s$ strange quark 
fugacity. In Table 1 we show the connection of the fugacities of the various 
particles we have included in our analysis with corresponding quark 
fugacities. 
 
\vspace{0.5cm} 
\begin{center} 
\begin{tabular}{|c|c|c||c|}\hline 
Hadrons, a & Quark structure & Fugacity, $\lambda_{\rm a}$ & \\ \hline\hline 
Light Unfl. Mesons & $q\bar{q}$ & $\lambda_q\lambda_q^{-1}=1$ & 
Non-strange \\ \cline{1-3} 
$N\;\&\;\Delta$ Baryons & $qqq$ & $\lambda_q^3 $ & 
Hadrons \\ \hline 
Kaons & $q\bar{s}$ & $\lambda_q\lambda_s^{-1}$ & 
\\ \cline{1-3} 
Hyperons ($\Lambda\;\&\;\Sigma$) & $qqs$ & $\lambda_q^2\lambda_s$ & 
Strange \\ \cline{1-3} 
$\Xi$ Baryons & $qss$ & $\lambda_q\lambda_s^2$ & 
Hadrons \\ \cline{1-3} 
$\Omega$ Baryons & $sss$ & $ \lambda_s^3$ & 
\\ \hline 
\end{tabular} 
 
\vspace{0.5cm} 
Table 1. Connection of the various hadrons with the corresponding quark 
fugacities. 
\end{center} 
\vspace{1cm} 
 
Since the u, d quarks carry baryon number 1/3 and strangeness number 0, 
whereas the corresponding number for the s quark are 1/3 and -1, 
respectively, we have the relations 
\[\hspace{5.7cm}\lambda_q=\lambda_B^{1/3}\;,\;\;\;\;\lambda_s= 
\lambda_B^{1/3}\lambda_S^{-1}\;,\hspace{4.8cm}(6a)\] 
or, equivalently, 
\[\hspace{6cm}\lambda_B=\lambda_q^3\;,\;\;\;\;\lambda_S= 
\lambda_q\lambda_s^{-1}\;.\hspace{5cm}(6b)\] 
 
We now commence with the main exposition of the present Section by 
displaying the generic form of the bootstrap equation in the presence of 
strangeness. It reads as follows (see also equ. (40) of Ref. [8]): 
\setcounter{equation}{6} 
\[\tilde{B}(p^2)\tilde{\tau}(p^2,b,s)= 
\underbrace{g_{bs}\tilde{B}(p^2)\delta_0(p^2-m_{bs}^2)}_{input\;term} 
+\sum_{n=2}^\infty\frac{1}{n!}\int\delta^4\left(p-\sum_{i=1}^np_i\right)
\cdot\]
\begin{equation} 
\cdot\sum_{\{b_i\}}\delta_K\left(b-\sum_{i=1}^nb_i\right) 
\sum_{\{s_i\}}\delta_K\left(s-\sum_{i=1}^ns_i\right) 
\prod_{i=1}^n\tilde{B}(p_i^2)\tilde{\tau}(p_i^2,b_i,s_i)d^4p_i\;, 
\end{equation} 
where the $g_{bs}$ represent degeneracy factors applicable to the given set 
of labels and where the tilde on $B$ and $\tau$ imply a rearrangement of the 
form 
\begin{equation} 
B(p^2)\tau(p^2,b,s)\equiv\tilde{B}(p^2)\tilde{\tau}(p^2,b,s) 
\end{equation} 
with respect to our definitions in the previous Section. 
 
We shall refer to the first term appearing on the right hand side of (7) as 
{\bf input term}, $\varphi$, and to the second as the {\bf mass spectrum 
containing} term, $G$. The specific choice one makes in using the above 
relation is of crucial significance as far as the dynamical description of the 
SBM is concerned. 
 
Next, we carry out three Laplace transformations, one continuous and two 
discrete, which lead to the replacements: 
\begin{equation} 
(p_\mu ,b,s)\rightarrow(\beta_\mu ,\lambda_B , \lambda_S)\;, 
\end{equation} 
with the dual variables $\beta_\mu,\;\lambda_B,\;\lambda_S$ 
corresponding, respectively, to inverse four-temperature, baryon and 
strangeness fugacities. 
Specifically, we have 
\[\varphi(\beta,\lambda_B,\lambda_S)=\sum_{b=-\infty}^{\infty} 
\lambda_B^b\sum_{s=-\infty}^{\infty}\lambda_S^s 
\int e^{-\beta^\mu p_\mu}g_{bs}\tilde{B}(p^2)\delta_0(p^2-m_{bs}^2) 
dp^4\] 
\begin{equation} 
\hspace{1cm} =\frac{2\pi}{\beta}\sum_{b=-\infty}^{\infty} 
\lambda_B^b\sum_{s=-\infty}^{\infty}\lambda_S^s 
g_{bs}\tilde{B}(m_{bs}^2) m_{bs}K_1(\beta m_{bs}) 
\end{equation} 
for the Laplace-transformed input term and 
\[G(\beta,\lambda_B,\lambda_S)=\sum_{b=-\infty}^{\infty} 
\lambda_B^b\sum_{s=-\infty}^{\infty}\lambda_S^s 
\int e^{-\beta^\mu p_\mu}\tilde{B}(p^2)\tilde{\tau}(p^2,b,s)dp^4\] 
\begin{equation} 
\hspace{1cm}=\frac{2\pi}{\beta}\int_0^{\infty} m\tilde{B}(m^2) \tilde{\tau}
(m^2,\lambda_B,\lambda_S)K_1(\beta m)dm^2 
\end{equation} 
for the Laplace-transformed mass-spectrum containing term. In the above 
relations $K$ denotes the modified Bessel function of the second kind. 
 
According to standard procedure [11] the bootstrap equation assumes the 
form 
\begin{equation} 
\varphi(\beta,\lambda_q,\lambda_s)=2G(\beta,\lambda_q,\lambda_s) 
-\exp [G(\beta,\lambda_q,\lambda_s)]+1 
\end{equation} 
and displays, in the $\varphi-G$ plane, a square root branch point at (see 
Fig. 1) 
\[\hspace{5.5cm}\varphi(T_{cr},\mu_{q\;cr},\mu_{s\;cr})=\ln 4-1\;, 
\hspace{5cm}(13a)\] 
\[\hspace{5.7cm} G(T_{cr},\mu_{q\;cr},\mu_{s\;cr})=\ln 2\;. 
\hspace{5.3cm}(13b)\] 
Eq. (13a) defines a {\bf critical surface} in the 3-d space $(T,\mu_q,\mu_s)$ 
which sets the limits of the hadronic matter. The region radially outside the 
critical surface belongs to unphysical solutions of the bootstrap equation and 
is thereby assigned to a new phase of matter, presumably the QGP phase. 
 
Let us also introduce the temperature $T_0$ according to 
\setcounter{equation}{13} 
\begin{equation}
\varphi(T_0,\lambda_q=1,\lambda_s=1)=\ln 4-1\;, 
\end{equation} 
which constitutes the highest temperature beyond which the Hadron Gas 
phase does not exist. 
 
Next, we turn our attention to the thermodynamical description of the 
system. According to the bootstrap scheme, the number of available states in 
a volume $d^3p$ around $\vec{p}$, baryon number $b$ and strangeness 
number $s$ is given, in covariant form, by 
\begin{equation} 
\frac{2V_{\mu}^{ext}p^{\mu}}{{(2\pi)}^3}\tilde{\tau}(p^2,b,s)d^4p\;, 
\end{equation} 
where $V_{\mu}^{ext}$ is the total external (four) volume available to the 
system. It is a constant as far as the integration over $d^4p$ is concerned. 
Accordingly, the grand canonical partition function for the system reads, in 
covariant form, 
\begin{equation} 
\ln\;Z(\beta,V,\lambda_B,\lambda_S)= 
\sum_{b=-\infty}^{\infty} 
\lambda_B^b\sum_{s=-\infty}^{\infty}\lambda_S^s 
\int \frac{2V_{\mu}p^{\mu}}{{(2\pi)}^3} 
\tilde{\tau}(p^2,b,s) e^{-\beta^\mu p_\mu} dp^4\;\;. 
\end{equation} 
Switching to quark fugacities, choosing the four-vectors $V^{\mu}$ and 
$\beta^\mu$ to be parallel and going to the frame for which 
$\beta^\mu=(\beta,0,0,0)$ [11], we write 
\begin{equation} 
\ln\;Z(\beta,V,\lambda_q,\lambda_s)= 
\frac{V}{\beta2\pi^2}\cdot\int_0^{\infty}m^2 
\tilde{\tau}(m^2,\lambda_q,\lambda_s) K_2(\beta m) dm^2\;\;. 
\end{equation} 
Our problem now is to express the above partition function in terms of the 
function $G(\beta,\lambda_q,\lambda_s)$ which contains the bootstrap mass 
spectrum. Once this is done we shall be in position to extract specific results 
from the extended SBM, via the inclusion of strangeness. This we shall do in 
the next two Sections by referring to specific versions of our extended SBM. 

We close the present Section with phenomenologically motivated remarks
which will set the tone for our subsequent applications. Let us return to 
equation (4). The volume to mass constant provides a quantity that can be 
related to the MIT bag model [9]. We set 
\begin{equation} 
\frac{V}{m}=\frac{V_i}{m_i}=\frac{1}{4B}\;\;\;, 
\end{equation} 
where $B$ is the MIT bag constant and where $V_i(m_i)$ denotes the 
volume (mass) of the fireball. The first equality in (18) comes from the 
assumption that the volume (mass) of a given fireball is the sum of the 
volumes (masses) of the constituent fireballs. 
 
Let us assess the splitting between the $B(p^2)$ and $\tau(p^2,b,s)$ in the 
bootstrap equation. We start with the ``natural'' definition of $B(p^2)$ as 
given by (5). Here, we have a purely kinematical assignment to this quantity, 
so all dynamics of the bootstrap model are carried by $\tau(p^2,b,s)$ [12]. 
Setting $B(m^2)\equiv H_0 m^2$ we find in this case 
\begin{equation} 
H_0=\frac{2}{(2\pi)^3 4B}\;\;\;. 
\end{equation} 
 
A rearrangement of the factors $\tilde{B}$ and $\tilde{\tau}$ would imply a 
behaviour of the form $\tilde{B}(m^2)=const \cdot m^d$. Any choice for 
which $d\neq 2$ entails an absorption of part of the dynamics into 
$\tilde{B}$. Traditionally, SBM applications have centered around the 
choice $\tilde{B}(m^2) \sim m^0$. Setting $B(m^2) \equiv H_2$ we are 
now obliged to introduce a reference mass scale $\tilde{m}$ in order to 
relate $\tilde{\tau}$ with $\tau$: 
\begin{equation} 
\tilde{\tau}(m^2,b,s)=\frac{m^2}{{\tilde{m}}^2}\tau(m^2,b,s)\;\;\;. 
\end{equation} 
We also determine, for this case, 
\begin{equation} 
H_2=\frac{2\tilde{m}^2}{(2\pi)^3 4B}\;\;\;. 
\end{equation} 
We stress that for any choice other than the one given by (19), one is forced 
to enter a reference mass scale into SBM descriptions. 

Given the above remarks, relevant to the phenomenological connection with
QCD, let us turn our attention to the asymptotic behaviour of the mass 
spectrum function $\rho(m)$, as $m\rightarrow\infty$. It can be shown 
[12,13,14] that 
\begin{equation} 
\tilde{B}(m^2)\tilde{\tau}(m^2,\{\lambda\})\stackrel{m\rightarrow\infty}
{\longrightarrow}C(\{\lambda\})m^{-3} \exp [m/T^*(\{\lambda\})]\;\;\;, 
\end{equation} 
where $T^*(\{\lambda\})$ satisfies the criticality equation, cf. Eq. (13a). In 
the above relation $\{\lambda\}$ is a collective index for fugacities, while 
$C(\{\lambda\})$ is a quantity independent of mass. 
 
For a given choice $\tilde{B}(m^2)=const\cdot m^d$ we have 
\[\tilde{\tau}(m^2,\{\lambda\})\stackrel{m\rightarrow\infty} 
{\longrightarrow} 
C'(\{\lambda\})m^{-3-d} \exp [m/T^*(\{\lambda\})]\;\;\;,\] 
or equivalently, with reference to (3), 
\begin{equation} 
\tilde{\rho}(m^2,\{\lambda\})\stackrel{m\rightarrow\infty} 
{\longrightarrow} 
2C'(\{\lambda\})m^{-\alpha} \exp [m/T^*(\{\lambda\})]\;\;\;, 
\end{equation} 
where $\alpha=2+d$. 
 
The choices, entailed by relations (19) and (21), correspond to $\alpha=4$ 
and $\alpha=2$, respectively. These two cases facilitate analytic procedures 
linking the grand canonical partition function to the term $G$, which 
contains the mass spectrum, and eventually, through the bootstrap equation, 
to the input term $\varphi$.

\vspace{2cm} 
{\large{\bf 3. Study of the {\boldmath $\alpha$}$\bf =2$
version of the SBM}}
 
In this section we shall study that version of the SBM which is dictated by 
the choice (21), equivalently $\alpha=2$. At the hadronic level the input 
function is furnished by 
\begin{equation} 
\varphi(T,\lambda_q,\lambda_s)=2\pi H_2 T 
\sum_{\rm a} (\lambda_{\rm a}+\lambda_{\rm a}^{-1})\sum_i g_{{\rm a}i}
 m_{{\rm a}i} K_1 (\frac{m_{{\rm a}i}}{T})\;\;, 
\end{equation} 
where the index ``${\rm a}$'' runs over all strange and non-strange hadrons.
Notice that on the left hand side we have entered fugacities in terms of quark 
quantum numbers. The translation has to be made hadron-by-hadron 
according to the specifications given in the previous Section, cf. equ (6b). 
 
The critical surface is determined by (13a), which represents a constraint 
among the co-ordinates in the 3-dimensional space 
$(T,\lambda_q,\lambda_s)$. The presence of the independent\footnote
{In the sense that, according to (21), it can be introduced
independently of the bag constant $B$ which corresponds to a second 
phenomenological input parameter for the model.} 
parameter $H_2$ in (24) allows us to use $T_0$ as an input to the model 
through the condition 
\begin{equation} 
\varphi(T_0,\lambda_q=1,\lambda_s=1\;;\;H_2)=\ln 4-1\;, 
\end{equation} 
which relates $H_2$ to $T_0$. One can now determine the critical surface 
numerically for each given choice of to $T_0$. The relevant study will be 
presented in Section 5 where we shall make an assessment of our results. 
 
Next, we turn our attention to the surface, in $(T,\lambda_q,\lambda_s)$ 
space, for which the expected value for the strangeness quantum number is 
zero. Our first task is to find a suitable expression for the grand partition 
function which relates it to bootstrap model quantities. 
 
With reference to [11] let us choose a frame for which our four-volume and 
four-temperature are parallel to each other. If we also take the viewpoint of 
that inertial observer for whom $T_\mu =(T,0,0,0)$, then we may write 
\begin{equation} 
\ln\;Z(\beta,V,\lambda_q,\lambda_s)= 
\frac{2V}{(2\pi)^3}\int p^0 e^{-\beta p_0} \tau_2 
(p^2, \lambda_q,\lambda_s) d^4p\;\;. 
\end{equation} 
We stress that in the above equation $V$ is {\bf not} the fireball volume, as 
is the case with (16), but the volume in which the thermodynamical system is 
enclosed. From (11) and (20) we have that 
\begin{equation} 
G(\beta,\lambda_q,\lambda_s)=H_2\int e^{-\beta p_0} \tau_2 
(p^2, \lambda_q,\lambda_s) d^4p\;\;. 
\end{equation} 
Putting the last two relations together we obtain 
\begin{equation} 
\ln\;Z(V,\beta,\lambda_q,\lambda_s)= 
-\frac{2V}{(2\pi)^3 H_2} \frac{\partial}{\partial \beta} 
G(\beta,\lambda_q,\lambda_s)\;\;. 
\end{equation} 
Referring to the bootstrap equation, (12), we easily determine 
\begin{equation} 
\frac{\partial G(\beta,\lambda_q,\lambda_s)}{\partial \beta}= 
\left (\frac{d\varphi}{dG} \right )^{-1} 
\frac{\partial \varphi(\beta,\lambda_q,\lambda_s)}{\partial \beta}= 
\frac{1}{2-e^{G(\beta,\lambda_q,\lambda_s)}}\cdot 
\frac{\partial \varphi(\beta,\lambda_q,\lambda_s)}{\partial \beta}\;\;. 
\end{equation} 
Therefore, 
\begin{equation} 
\ln\;Z(V,\beta,\lambda_q,\lambda_s)= 
-\frac{2V}{(2\pi)^3 H_2} \cdot 
\frac{1}{2-e^{G(\beta,\lambda_q,\lambda_s)}} 
\frac{\partial \varphi(\beta,\lambda_q,\lambda_s)}{\partial \beta}\;\;. 
\end{equation} 

We remark that the above equation refers to point particles. A realistic
description of the system calls for volume corrections [15] which bring the 
bag constant $B$ as an explicit, new input parameter. For our purposes, this 
procedure can be bypassed as we shall argue, {\it a posteriori}, shortly. 
 
The zero strangeness condition is given by 
\begin{equation} 
\lambda_s \left.\frac{\partial \ln\;Z(V,\beta,\lambda_q,\lambda_s)} 
{\partial \lambda_s} \right|_{( V,\beta,\lambda_q)} =0 
\end{equation} 
which, with the aid of (30), becomes 
\begin{equation} 
\frac{\partial}{\partial \lambda_s} 
\left [ \frac{1}{2- \exp G(\beta,\lambda_q,\lambda_s)} \right ] 
\frac{\partial \varphi(\beta,\lambda_q,\lambda_s)}{\partial \beta}+ 
\frac{1}{2- \exp G(\beta,\lambda_q,\lambda_s)} 
\frac{\partial^2 \varphi(\beta,\lambda_q,\lambda_s)} 
{\partial \lambda_s \partial \beta}\;=0 
\end{equation} 
and finally 
\begin{equation} 
\frac{e^{G(\beta,\lambda_q,\lambda_s)}} 
{\left [ 2- \exp G(\beta,\lambda_q,\lambda_s) \right] ^2} \cdot 
\frac{\partial \varphi(\beta,\lambda_q,\lambda_s)}{\partial \lambda_s}\cdot 
\frac{\partial \varphi(\beta,\lambda_q,\lambda_s)}{\partial \beta}+ 
\frac{\partial^2 \varphi(\beta,\lambda_q,\lambda_s)} 
{\partial \lambda_s \partial \beta}\;=0 
\end{equation} 
 
The above relation specifies a surface in the three-dimensional space of 
$(T,\lambda_q,\lambda_s)$ onto which a hadronic system of zero 
strangeness must find itself. We immediately notice that the $<S>=0$ 
surface ``knows'' when it is approaching the critical point, i.e. a point on the 
critical surface since, for $G=\ln 2$, the denominator in the first factor on 
the rhs of (33) vanishes. This implies that the values of 
$(T,\lambda_q,\lambda_s)$ which lead the system to criticality are 
imprinted on the surface of zero strangeness. 
 
A second characteristic of the $<S>=0$ contour is the fact that both strange 
and {\bf non-strange} particles contribute to its specification. The presence 
of the non-strange particles, surviving the partial differentiation with respect 
to $\lambda_s$, reside in the quantities $G(\beta,\lambda_q,\lambda_s)$ and 
$\partial \varphi(\beta,\lambda_q,\lambda_s) / \partial \beta$. This is a 
significant fact, given our expectation that the critical temperature $T_0$, 
corresponding to $\lambda_s =1$, is of the order of the pion mass. 
 
We finally remark that if, in the place of total strangeness, we had based our 
considerations on strangeness density, then we would have found 
\begin{equation} 
<s>=\frac{<S>}{<V>}= 
\frac{<S>}{\Delta \left( 1+ \frac{\textstyle \varepsilon_{pt} 
(\beta,\lambda_q,\lambda_s)}{\textstyle 4B} \right )}\;\;, 
\end{equation} 
where we have taken into account the volume correction factor [15]. 
Accordingly, the imposition of zero strangeness density, which is enforced 
by the vanishing of the numerator, would lead back to (33) and our analysis 
would stand in its present form. 
 
Our final undertaking of this Section is the confrontation of the problem 
regarding the intersection of zero-strangeness surface with the critical one, 
given the singular behaviour exhibited by the former on account of the 
condition $G=\ln 2$. Assuming that (33) continues to hold true on the 
critical surface and given that the quantity 
$\partial^2 \varphi(\beta,\lambda_q,\lambda_s) / \partial \beta \partial 
\lambda_s$ 
does not exhibit any singularity at the critical point we are led to impose 
the condition 
\begin{equation} 
\exp \left[ G(\beta,\lambda_q,\lambda_s) \right]\cdot 
\frac{\partial \varphi(\beta,\lambda_q,\lambda_s)}{\partial \lambda_s}\cdot 
\frac{\partial \varphi(\beta,\lambda_q,\lambda_s)}{\partial \beta} 
\;\;\rightarrow\;\;0\;\;, 
\end{equation} 
which, in turn, can only be satisfied if 
\begin{equation} 
\left.\frac{\partial \varphi(\beta_{cr},\lambda_{q\;cr},\lambda_s)} 
{\partial \lambda_s} \right| _{\lambda_s=\lambda_{s\;cr}}\;=\;0\;\;. 
\end{equation} 
The above and eq. (13a) for the critical surface comprise a 
system of two equations that can be numerically solved and hence lead to the 
curve which represents the intersection of the two surfaces of interest. 
 
The results of our numerical study of the $<S>=0$ surface and of its 
approach to the critical surface will be presented in Section 5. The next issue 
to occupy our attention is the consideration of the $\alpha=4$ version of the 
strangeness incorporating SBM.

\vspace{2cm} 
{\large{\bf 4. Study of the {\boldmath $\alpha$}$\bf =4$
version of the SBM}}
 
The $\alpha=4$ version of the SBM represents the first, integer-valued case 
for which the asymptotic exponent $\alpha$ is larger than 7/2, namely the 
critical value which, in the absence of volume corrections, allows for the 
existence of a new phase of matter beyond the Hagedorn temperature [7]. 
Our relevant study will follow the patterns established in the previous 
Section for the $\alpha=2$ version of the model. 
 
The input term assumes the form 
 \begin{equation} 
\varphi(T,\lambda_q,\lambda_s)=2\pi H_0 T 
\sum_{\rm a} (\lambda_{\rm a}+\lambda_{\rm a}^{-1})\sum_i g_{{\rm a}i}
m_{{\rm a}i}^3 K_1 (\frac{m_{{\rm a}i}}{T})\;\;, 
\end{equation} 
where, now, the parameter $H_0$ is directly related to the bag constant, cf. 
eq. (19). So either of the two can be considered as independent but not both. 
In this case, we have a direct correspondence between $T_0$ and $B$. The 
critical surface can now be (numerically) determined in the 
$(T,\lambda_q,\lambda_s)$ space, once the fugacities $\lambda_{\rm a}$ are 
adjusted to quark-related values. Results of the relevant study will be 
presented and discussed in the following Section. 
 
We now proceed to consider the zero strangeness surface. We start by 
determining the form acquired by the bootstrap function $G$ for the case in 
hand. We find 
\[G(\beta,\lambda_q,\lambda_s)=\frac{2\pi}{\beta} 
\int_0^{\infty} m B_0 (m^2)\tau_0(m^2, \lambda_q,\lambda_s) 
K_1(\beta m) dm^2\] 
\begin{equation} 
\hspace{1cm} =\frac{2\pi H_0}{\beta}\int_0^{\infty} 
m^3 \tau_0(m^2, \lambda_q,\lambda_s) K_1(\beta m) dm^2\;\;\;. 
\end{equation} 
We shall use the relation 
\begin{equation} 
\frac{d}{d\beta} [\beta ^2 K_2 (m \beta)]=-m\beta^2 K_1(m\beta)\;\;, 
\end{equation} 
which follows from the Bessel function identities 
$K_{n+1}(x)=\left(\frac{2n}{x}\right) K_n(x)+ K_{n-1}(x)$ and 
$K'_n(x)=-\frac{1}{2}[K_{n+1}(x)+ K_{n-1}(x)]$. Further, taking into 
account the asymptotic property 
\begin{equation} 
\lim_{x\rightarrow\infty}x^2 K_2(mx)= 
\lim_{x\rightarrow\infty}\sqrt{\frac{\pi}{2}}x^{3/2}e^{-x}\;=\;0 
\end{equation} 
the logarithm of the grand canonical partition function (17) becomes 
\begin{equation} 
\ln\;Z(\beta,V,\lambda_q,\lambda_s)= 
\frac{V}{\beta^3 2\pi^2}\cdot\int_0^{\infty}m^2 
\tau_0(m^2,\lambda_q,\lambda_s) 
\left\{\int_{\infty}^{\beta}\frac{d}{dx} [x^2 K_2(xm)]dx \right\}dm^2\;\;. 
\end{equation} 
Use of (39) and (38) finally gives 
\begin{equation} 
\ln\;Z(\beta,V,\lambda_q,\lambda_s)= 
\frac{V}{4\pi^3 H_0}\frac{1}{\beta^3}\int_{\beta}^{\infty} 
x^3 G(x,\lambda_q,\lambda_s)dx\;\;. 
\end{equation} 
The zero strangeness condition now translates into 
\begin{equation} 
\int_{\beta}^{\infty} x^3 \frac{\partial G(x,\lambda_q,\lambda_s)} 
{\partial \lambda_s}dx\;=\;0\;\;. 
\end{equation} 
With the aid of the bootstrap equation, relation (43) becomes 
\begin{equation} 
\int_{\beta}^{\infty} x^3 \frac{1}{2-\exp [G(x,\lambda_q,\lambda_s)]} 
\frac{\partial \varphi (x,\lambda_q,\lambda_s)}{\partial \lambda_s} 
dx\;=\;0\;\;. 
\end{equation} 
 
Making the variable change x=1/y, the above relation takes the form 
\begin{equation} 
\int_0^T \frac{1}{y^5} \frac{1}{2-\exp [G(y,\lambda_q,\lambda_s)]} 
\frac{\partial \varphi (y,\lambda_q,\lambda_s)}{\partial \lambda_s} 
dy\;=\;0\;\;, 
\end{equation} 
which is well defined, having no problems at $y=0$ despite the presence of 
the factor $y^{-5}$. The point is that $\varphi (y,\lambda_q,\lambda_s)$ 
carries with it terms containing the Bessel function $K_1 (m/y)$ which is 
not affected by the partial differentiation with respect to $\lambda_s$. But, 
the argument of the $K_1$ functions tends to infinity as $y\rightarrow 0$, 
which leads to $\lim_{y\rightarrow 0} y^{-5} K_1(m/y)=0$. On the other 
hand, the problem associated with the approach to the critical point, at 
$G=\ln 2$, persists as with the $\alpha=2$ version of the previous Section. 
So, even if the integral is well defined, we shall have to face tedious 
convergence problems during the course of a numerical analysis which aims 
at determining the intersection of the $<S>=0$ with the critical surface. 
 
We bypass the aforementioned problem by introducing a new variable, as 
follows. Setting 
\begin{equation} 
z=2-\exp [G(y,\lambda_q,\lambda_s)]\;\;, 
\end{equation} 
we determine 
\begin{equation} 
dz=-\frac{e^G}{2-e^G}\frac{\partial \varphi (y,\lambda_q,\lambda_s)} 
{\partial y} dy\;\;. 
\end{equation} 
From (46) it also follows that $-e^G =z-2$, whereupon we conclude 
\begin{equation} 
\frac{dy}{2-\exp G(\beta,\lambda_q,\lambda_s)}= 
\frac{dz}{z-2}\left.\left\{\left[ 
\frac{\partial \varphi (y,\lambda_q,\lambda_s)}{\partial y} 
\right]^{-1}\right\} \right|_ 
{\textstyle z=2-\exp [G(y,\lambda_{q\;cr},\lambda_{s\;cr})]}. 
\end{equation} 
 
Next, we must determine the limits for the dz integration. We assume that 
the fugacities $\lambda_q$ and $\lambda_s$ remain finite as $y\rightarrow 
0$. But then, the bootstrap equation, along with the fact that 
$\lim_{y\rightarrow 0} 
\varphi(y,\lambda_q,\lambda_s)=0$, 
gives $z=2-e^0=1$. For the upper limit, the real interest lies with T 
approaching a critical 
value. At this point we have $z=2-e^{\ln 2}=0$. We, therefore, obtain 
\begin{equation} 
\int_1^0\frac{dz}{z-2}\cdot\left[ 
\frac{\frac{\textstyle \partial \varphi (y,\lambda_{q\;cr},\lambda_{s\;cr})} 
{\textstyle \partial \lambda_s}} 
{y^5 \cdot\frac{\textstyle 
\partial \varphi (y,\lambda_{q\;cr},\lambda_{s\;cr})} 
{\textstyle \partial y}} \right]_ 
{\textstyle z=2-\exp [G(y,\lambda_{q\;cr},\lambda_{s\;cr})]}\;\;=\;0\;\;. 
\end{equation} 
Notice that at the critical point, where $z=0$, no singularity appears, which 
explicitly shows that the (apparent) singularity at $G=\ln 2$ is integrable. 
 
As it turns out, the z-variable is favoured with respect to convergence 
properties at the upper limit, i.e when one is approaching a critical 
surface value, whereas the y-integration is more efficiently performed at the 
lower limit. The best of both cases is attained by introducing an intermediate 
temperature $\tilde{T}$, conveniently chosen so that the function which is 
integrated vanishes at this temperature, and determine $<S>=0$ surface via 
the equation 
\newpage 
\[\int_0^{\tilde{T}} 
\frac{1}{2-\exp [G(y,\lambda_{q\;cr},\lambda_{s\;cr})]} 
\frac{\partial \varphi (y,\lambda_{q\;cr},\lambda_{s\;cr})} 
{\partial \lambda_s} 
\frac{dy}{y^5}+\hspace{4cm}\] 
\begin{equation} 
\int_{\tilde{z}}^0\frac{dz}{z-2}\cdot\left[ 
\frac{\frac{\textstyle \partial \varphi (y,\lambda_{q\;cr},\lambda_{s\;cr})} 
{\textstyle \partial \lambda_s}} 
{y^5 \cdot\frac{\textstyle \partial \varphi 
(y,\lambda_{q\;cr},\lambda_{s\;cr})} 
{\textstyle \partial y}} \right]_ 
{\textstyle z=2-\exp [G(y,\lambda_{q\;cr},\lambda_{s\;cr})]}\;\;=\;0\;\;. 
\end{equation} 
Our specification for $\tilde{T}$ is fulfilled if it is determined by the 
relation 
\begin{equation} 
\left.\frac{\partial \varphi (\tilde{T},\lambda_{q\;cr},\lambda_s)} 
{\partial \lambda_s}\right|_{\lambda_s=\lambda_{s\;cr}}\;\;=\;0 
\end{equation} 
according to which the lower limit of the $dz$-integration in (50) is 
\begin{equation} 
\tilde{z}=2-\exp [G(\tilde{T},\lambda_{q\;cr},\lambda_{s\;cr})]\;\;. 
\end{equation} 
 
In the next Section we shall present the results of our numerical investigation 
of the $\alpha=4$ version of the SBM and argue in favour of its suitability 
over the $\alpha=2$ version, as far as its implications near the critical 
surface are concerned.

\vspace{2cm} 
{\large{\bf 5. Behaviour near the Critical Surface}} 
 
In this Section we present the results of our numerical calculations of the 
critical and zero-strangeness surfaces in the $(T,\mu_q,\mu_s)$
three-dimensional space of the two versions of the SBM discussed in the previous 
sections. Specifically, we shall study the corresponding behaviours of the 
$\alpha=2$ and $\alpha=4$ cases as one approaches the critical surface, 
which, in the SBM context, signifies the termination of the hadronic world. 
Notice that we have chosen chemical potential variables in terms of which to 
present our findings, instead of fugacities.

\vspace{0.5cm} 
{\large{\bf 5.1. {\boldmath $\alpha$}$\bf =2$ case}} 
 
Let us begin with the critical surface for the case of $\alpha=2$. Choosing as 
input value 180 MeV for the $T_0$ variable, we depict, in figures 2a,b, two 
different profiles of the aforementioned surface. Fig. 2a shows that for a 
range of values of $\mu_s$ in the interval [0, 100] MeV, the critical surface 
is virtually perpendicular to the $T-\mu_q$ plane. This indicates that, for a 
system in this range of $\mu_s$ values, the critical surface is basically 
determined by $T$ and $\mu_q$. One also notices that the curves 
corresponding to constant $\mu_s$ converge to a value of about 330 MeV 
for $\mu_q$, as $T\rightarrow 0$, whilst (Fig. 2b) curves of constant 
$\mu_q$, in the same limit and on the positive side, towards 
$\mu_s\simeq 560$ MeV. 
 
The dependence of the critical surface on the input parameter $T_0$, is the 
object of attention in figures 3a,b. We depict characteristic intersections of 
the $\mu_s=0$ and $\mu_q=0$ planes, respectively, by the critical surface, 
for different values of $T_0$ . We notice that towards low temperatures the 
dependence on $T_0$ is very slight. One must keep in mind, however, that 
this is the least trustworthy region of our analysis, since the Bolztmann 
approximation is rather inadequate for low temperatures. 
 
Turning our attention to the $<S>=0$ surface we choose a fixed value for 
the ratio $\mu_q/T$, 0.4 to be specific, and study the variation of $\mu_s$ 
with temperature. The result of the relevant numerical study is shown in 
figure 4 for various values of $T_0$. The vertical lines in the figure show 
the edge of the cylindrical, critical surface corresponding to the same value 
of the $\mu_q/T$ ratio, in the vicinity of small values for the $\mu_s$ 
chemical potential. Finally, the uninterrupted curve represents the zero 
strangeness contour, for $\mu_q/T=0.4$, which results from an ``ideal hadron 
gas'' analysis [16,17] wherein the bootstrap equation does not provide an 
input to the partition function. It is satisfying to witness an agreement 
between the two descriptions up to a certain point. The observed tendency is 
for the bootstrap description to drive towards higher values for $\mu_s$ as 
the critical surface is approached\footnote{In the ``ideal hadron gas'' model, 
of course, there is no critical surface.}. At a given point the chemical 
potential $\mu_s$ reverses its course and merges with the critical surface on 
its way up. Finally, in the same figure we exhibit an extreme case, 
corresponding to $T_0=250$ MeV, for which the $<S>=0$ surface dips into 
the negative value region for the strange chemical potential before merging 
with the critical surface. 
 
In figures 5a,b we depict the profile of the intersection curve between the 
$<S>=0$ and critical surfaces for $\alpha=2$, by projecting it on the 
$(\mu_{q\;cr},\mu_{s\;cr})$ and $(T,\mu_{s\;cr})$ planes, respectively, for 
various values of $T_0$. Note that there exists a value for $T_0$, equal to 
about 232 MeV, for which the intersection curve stays for long distance in 
touch with $\mu_{s\;cr}=0$. For smaller values of $T_0$, the 
$(\mu_{q\;cr},\mu_{s\;cr})$ dependence along the intersection curve is 
monotonous, while below the aforementioned value, it dips into the negative 
$\mu_{s\;cr}$ range before rising up.

\vspace{0.5cm} 
{\large{\bf 5.2. {\boldmath $\alpha$}$\bf =4$ case}} 
 
We now turn our attention to the $\alpha=4$ version of the SBM which has 
received minimal attention in the past. The present incorporation of 
strangeness into the bootstrap scheme will bring out decisive advantages of 
this over the $\alpha=2$ case, which has dominated SBM studies up to now. 
 
The numerical study of the $\alpha=4$ case is quite demanding, certainly 
more so than its $\alpha=2$ counterpart. It might be helpful to briefly 
remark on the procedure employed for the determination of the $<S>=0$ 
surface, for $\alpha=4$. Starting from (45) we give input values to the 
quantities $(\lambda_q,T)$ and compute the $\lambda_s$ value for which 
the left hand side vanishes. We use the Newton-Raphson method which also 
requires knowledge of the first derivative of the function. The latter is given 
by 
\[\int_0^T\frac{dy}{y^5}\left\{ 
\frac{\exp [G(y,\lambda_q,\lambda_s)]} 
{\{2-\exp [G(y,\lambda_q,\lambda_s)]\}^3} 
\left[\frac{\partial \varphi (y,\lambda_q,\lambda_s)}{\partial \lambda_s} 
\right]^2 \right. +\hspace{3cm}\] 
\begin{equation} 
+\left.\frac{1}{2-\exp [G(y,\lambda_q,\lambda_s)]} 
\frac{\partial^2 \varphi (y,\lambda_q,\lambda_s)}{\partial \lambda_s^2} 
\right\}\;=\;0\;\;. 
\end{equation} 
The integration in (45) is performed via appropriate routines. 
 
For temperatures which approach the critical one, i.e. 
$T^*(\lambda_{q\;cr},\lambda_{s\;cr})$, computations are facilitated by 
employing the transformation $w=(T^*-y)^{1/2}$. To see this observe that 
the encountered singularity is of the form $(2-e^G)^{-1}$, equivalently, via 
the use of the bootstrap equation, $\frac{1}{2}(\varphi_0-\varphi)^{-1/2}$. 
But, then, a Taylor expansion of $\varphi$ around $T^*$, 
\begin{equation} 
\varphi_0 -\varphi(y,\lambda_q,\lambda_s)\approx 
C[T^*(\lambda_{q\;cr},\lambda_{s\;cr})] 
[T^*(\lambda_{q\;cr},\lambda_{s\;cr})-y]\;\;, 
\end{equation} 
ascertains that the aforementioned transformation does away with the 
singularity. 
 
Not knowing $T^*$ {\it a priori}, use of the $w$ variable gives satisfactory 
results for values of $T$ in the vicinity of $T*$ but not for $T=T^*$. The 
problem is confronted by appealing to (50). Using as input a critical value 
for the up-down quark fugacity, $\lambda_q=\lambda_{q\;cr}$, we 
determine the corresponding value for $\lambda_{s\;cr }$. We do not use 
the Newton-Raphson method for extracting $\lambda_{q\;cr }$ given the 
complex expression entering the left hand side of (50). Instead, we employ 
the ``Brent'' method which furnishes a value for the root once we specify an 
interval wherein the said root lies. Finally, $T^*$ is determined via a 
numerical solution of the equation $\varphi (T^*,\lambda_q,\lambda_s)=\ln 
4- 
1$. 
 
Figures 6-10 summarise the findings of our numerical investigations for the 
$\alpha=4$ version of the extended SBM. In figures 6a,b we depict the 
profile of the critical surface in a similar manner as done for the $\alpha=2$ 
case in figures 2a,b. The critical surface does not depend appreciably on 
$\mu_s$ for an extended range of values on either side of the $\mu_s=0$ 
(Fig. 6b illustrates this occurrence only for positive values of $\mu_q$). 
Near $T=0$ the critical surface converges towards $\mu_q\approx 325$ 
MeV for $\mu_s=0$ (Fig. 6a) and $\mu_s\approx 550$ MeV for $\mu_q=0$ 
(Fig. 6b). In figures 7a,b we exhibit intersections of the critical surface for 
different values of $T_0$, just as figures 3a,b show for the $\alpha=2$ case. 
 
A comparative study between the $\alpha=2$ and $\alpha=4$ versions of the 
SBM is presented in figures 8a,b. Here, for the fixed value of 180 MeV for 
$T_0$, we show the intersections of the critical surface with the planes 
$(T,\mu_q)$ and $(T,\mu_s)$, respectively, as well as a second, parallel 
intersection. One witnesses a similar behaviour of the two cases. 
 
The numerical study of the $<S>=0$ surface for the SBM choice 
$\alpha=4$, though much harder, {\it can} be achieved. As with the 
$\alpha=2$ case, we present, in figure 9, curves on the zero strangeness 
surface displaying the $T-\mu_s$ dependence for the constant ratio 
$\mu_q/T=0.4$ and for different values of $T_0$. As before, the (almost) 
vertical lines represent the critical surface and adhere to the same $\mu_q/T$ 
ratio. We note that the curves $\mu_s(T)$ on $<S>=0$ approach 
tangentially the critical surface, without reversing their course. Notice the 
remarkable coincidence with the ``ideal hadron gas'' model curve up to a 
temperature which is only 25 MeV, or so, away from the critical surface. 
 
Figures 10a,b are analogous to figures 5a,b, respectively, for the $\alpha=2$ 
case. We observe a similar behaviour of the intersection of the surface 
$<S>=0$ with the critical surface. But, note that the value of $T_0$ for 
which the intersection curve stays for long distance in touch with 
$\mu_{s\;cr}=0$, is now much smaller (and more realistic) and equals 183 
MeV. Finally, in figure 11 we give a three-dimensional plot in the 
$(T,\mu_q,\mu_s)$ space for $T_0=183$ MeV. We display different cuts of 
the surface of zero strangeness, which correspond to constant fugacity 
$\lambda_q$, fixed by the ratio $\mu_q/T$. On the same plot 
we also present the intersection of the zero-strangeness surface with the 
critical surface.

\vspace{0.5cm} 
{\large{\bf 5.3. Relative Assessments - Prevalence of the
{\boldmath $\alpha$}$\bf =4$ version}} 
 
Let us proceed with an evaluation of our present findings, starting with the 
$\alpha=2$ version of the strangeness including SBM. The sudden upward 
trend of the chemical potential $\mu_s$ is an unwelcome occurrence, despite 
the agreement with the ``ideal hadron gas'' analysis, prior to approaching 
criticality. Indeed, our expectation that a QGP state of matter lying beyond 
the critical point would imply a tendency of $\mu_s$ towards zero and not 
away from it. Even if there exists an intermediate phase between the 
hadronic and the QGP states, which interpolates between constituent and 
current quarks, one would still expect $\mu_s$ in this phase to be smaller 
than in the pure hadronic one. Therefore we fully expect $\mu_s$ to {\bf 
converge} towards zero as criticality is approached and not to move away 
from it. 
 
A second point of discouragement has to do with particle ratios, for which a 
more detailed discussion will be conducted in the next section. Let us take, 
for example, the $\bar{\Xi}/\Xi$ ratio. In the ``ideal hadron gas'' model it is a 
genuinely monotonic increasing function of temperature. This is a trend we 
expect to be enhanced, as the (assumed) QGP phase is reached. 
Unfortunately, within the framework of the $\alpha=2$ version of the SBM 
we find this ratio to go down with temperature, as the critical point is 
approached. 
 
On a purely theoretical basis it is especially bothersome that volume 
correction ramifications seem to have no impact on our $\alpha=2$
analysis\footnote{Indeed, as we have seen, it made no difference whether our 
description was based on strangeness, or strangeness density. Equivalently, 
the bag constant $B$ which enters finite volume corrections did not figure 
into our final results.}. 
Given that such corrections are integrally connected with the realisability of 
a state of matter beyond the hadronic one, for $\alpha\leq7/2$, our results 
seem to open anew the issue whether this particular value for $\alpha$ is 
compatible with the existence of a new phase beyond the Hagedorn 
temperature. Secondly, according to Letessier and Tounsi [14] the anomaly 
of the Bootstrap model, signifying the termination of the hadronic world, 
persists in the thermodynamic limit and is, therefore, genuinely connected 
with a phase transition only for $\alpha>5/2$. Once again, the $\alpha=2$ 
version of the SBM fails to meet the desired criterion. 
 
It is clear that all the above objections, whether on the phenomenological or 
the purely theoretical front, are automatically removed for the $\alpha=4$ 
version of the SBM. Moreover, for this case we have the strong consequence 
that $T_0$, the critical temperature for zero chemical potentials, is in a one- 
to-one correspondence with the bag constant $B$. Here we have the 
intriguing possibility of cross checking QCD lattice input, which aims at the 
determination of $T_0$ on the one hand and phenomenological information 
related to the bag constant $B$ on the other. In Table 2 we exhibit the 
connection between $T_0$ and $B$ values, as provided by the $\alpha=4$ 
SBM\footnote{From this point on, numerical assignments to physical quantities 
such as $T_0$, $B$ etc. employ updated results [18] with respect to what 
was used in [9].}. 
 
To summarise, the $\alpha=4$ case represents, in all respects, a self 
consistent model which seems best suited to approach the problem of 
(possible) QGP formation in relativistic, heavy nucleus-nucleus collisions 
from the hadronic side. On this basis we shall take up in the next Section 
numerical determinations of strange particle productions ratios which are 
directly relevant to experimental observation. 
 
\vspace{0.5cm} 
\begin{center} 
\begin{tabular}{|c|c|}\hline 
$T_0$ (MeV) & $B^{1/4}$ (MeV) \\ \hline\hline 
150&155.494 \\ \hline 
160&178.128 \\ \hline 
170&202.419 \\ \hline 
180&228.231 \\ \hline 
190&255.424 \\ \hline 
200&283.859 \\ \hline 
\end{tabular} 
 
\vspace{0.5cm} 
Table 2. Connection between $T_0$ and $B$. 
\end{center}

\vspace{2cm} 
{\large{\bf 6. Study of (strange) Particle Ratios through
{\boldmath $\alpha$}$\bf =4$ SBM version}} 
 
Particle ratios for given pairs of hadronic species, strange in particular, 
emerging from a heavy nucleus-nucleus collision region, constitutes an 
accessible experimental measurement for which a data base is readily 
available [18]. In the framework of our thermal description of the multi-particle 
system generated via the aforementioned type of collisions such particle 
ratios can be computed, as functions of $T$, $\mu_q$ and $\mu_s$, from 
our partition function. A pioneering work, in this connection and within 
framework of what we have been referring to as the ``ideal hadron gas 
model'', has been performed by Cleymans and Satz [19]. Recall that the ideal 
hadron gas model furnishes a thermal description of the multi-hadron system 
without SBM input. In what follows we shall adjust the theoretical 
determination of particle ratios to SBM constraints. Clearly, their evaluation 
at the critical surface, for different values of $T_0$, is directly pertinent to 
multi-particle production from the critical system. 
 
Given our reasons, in favour of the $\alpha=4$ version of the SBM, our 
considerations will be restricted to this case. A quintessential ingredient of 
our intended analysis is the inclusion of fugacities corresponding to {\bf 
hadron species} [20] in terms of which particle ratios should be given. Their 
introduction into our formalism enters at the level of equations (10), (11) 
and (16) for the $\varphi$ and $G$ terms of the bootstrap equations and the 
partition functions, respectively. In this way, we produce generalisations of 
the form, e.g. in the partition function case, 
\begin{equation} 
Z(V,\beta,\lambda_q,\lambda_s)\;\;\rightarrow\;\; 
Z(V,\beta,\lambda_q,\lambda_s,\ldots,\lambda_i,\ldots)\;\;, 
\end{equation} 
where $\lambda_i$ represents the fugacity of the {\it i}th hadron. 
 
In our thermal description the total number of the particle {\it i} is given by 
\begin{equation} 
N_i^{th}=\left.\left(\lambda_i\frac 
{\partial \ln Z(V,\beta,\lambda_q,\lambda_s,\ldots,\lambda_i,\ldots)} 
{\partial \lambda_i}\right)\right|_{\ldots=\lambda_i=\ldots=1} \;\;, 
\end{equation} 
where ``{\it th}'' stands for thermal. 
 
Referring to (42) we write, for the $\alpha=4$ version of the SBM, 
\begin{equation} 
N_i^{th}(V,T, \lambda_q,\lambda_s)=\frac{VT^3}{4\pi^3 H_0} 
\int_0^T \frac{1}{y^5} \frac{1}{2-\exp [G(y,\lambda_q,\lambda_s)]} 
\left. 
\frac{\partial \varphi (y,\lambda_q,\lambda_s,\ldots,\lambda_i,\ldots)} 
{\partial \lambda_i}\right|_{\ldots=\lambda_i=\ldots=1}\;dy. 
\end{equation} 
 
Our approach to the data should also take into account the fact that 
experiments measure, as a rule, the products of resonance decays. 
Accordingly, we shall refer to the number of particles which also includes 
secondary production processes. We thereby deal with particle numbers 
$N_i$ which, for a given species, have the form 
\begin{equation} 
N_i(V,T, \lambda_q,\lambda_s)=N_i^{th}(V,T, \lambda_q,\lambda_s)+ 
\sum_j b_{ij} N_j^{th}(V,T, \lambda_q,\lambda_s)\;\;, 
\end{equation} 
where the $b_{ij}$ are branching ratios corresponding to the decay of 
resonance {\it j} into the observed hadron species labelled by {\it i}. 
 
Switching to particle densities, since as the volume drops out in the ratio the 
volume corrections become irrelevant, equations (57) and (58) determine 
quantities of the form ($i\neq j$) 
\begin{equation} 
\frac{n_i}{n_j}(T, \lambda_q,\lambda_s)\equiv X_{ij} 
(T,\lambda_q,\lambda_s)\;\;. 
\end{equation} 
 
Imposition of the condition $<S>=0$ yields equation (45), which furnishes a 
constraint among the thermodynamical variables $T$, $\lambda_q$ and 
$\lambda_s$ and enables us to eliminate one of them in (59). 
 
In figures 12a-12i we depict various ratios involving at least one strange 
particle, for the $\alpha=4$ version of the SBM, as functions of the 
temperature $T$ for constant fugacity $\lambda_q$ ($\mu_q/T$ fixed at 0.4 
in these examples) and for different values of $T_0$. In each case we 
present the curve stemming from the ideal hadron gas model, for comparison 
purposes. 
 
We next turn our attention to critical values of particle ratios, namely ratios 
evaluated on the intersection between the critical and the $<S>=0$ surfaces. 
Our study will be conducted on the basis of adopting the critical temperature 
$T_{cr}$ as our independent variable. The results of our numerical analyses 
for different values of $T_0$ are depicted in figures 13a-13i. 
 
The issue of greatest interest is an assessment of how close, according to our 
theoretical description, are current experimental data to the border of the 
hadronic world. To this end, let us remove the criticality constraint and test 
the proximity of actual data to it. Fixing the theoretical quantity (59) to its 
experimental value $x_{ij}^{exp}$, as recorded in observations from 
nucleus-nucleus collisions, we set 
\begin{equation} 
\frac{n_i}{n_j}(T, \lambda_q,\lambda_s)= 
x_{ij}^{exp}\pm \delta x_{ij}^{exp}\;\;\;, 
\end{equation} 
where $\delta x_{ij}^{exp}$ is the experimental error in the measurement of 
the quantity $x_{ij}^{exp}$. With the use of (45) we can eliminate one of 
the three thermodynamical quantities, e.g. $\lambda_s$, and arrive at two 
equations of the form 
\begin{equation} 
\frac{n_i}{n_j}(T, \lambda_q)=x_{ij}^{exp}-\delta x_{ij}^{exp} 
\hspace{1cm},\hspace{1cm} 
\frac{n_i}{n_j}(T, \lambda_q)=x_{ij}^{exp}+\delta x_{ij}^{exp}\;\;\;. 
\end{equation} 
The above equations define a band of values in the $(T,\mu_q)$ plane which 
are compatible with (60). Consider now a second particle ratio, say 
$\frac{\textstyle n_k}{\textstyle n_l}$
($k\neq l$, $k\neq i$ and/or $j\neq l$). Repeating the
same procedure we determine a second band in the $(T,\mu_q)$ plane. If the 
region of intersection of two bands has a part {\bf inside} the region 
bounded by the critical surface (Fig. 14a) then, according to our theoretical 
proposal, the experimentally recorded ratios are compatible with 
multiparticle production within the hadronic phase. If, on the other hand, the 
intersection of the two bands occurs completely {\bf outside} the critical 
surface (Fig. 14b), then the situation merits further consideration. Indeed, 
one could consider the possibility that emission has occured from an 
equilibrated QGP state before the latter has reached the critical point and 
subsequently the content of the system as a whole transformed into hadrons 
which equilibrated and immediately afterwards freesed-out. Clearly, all this 
is based on our assumption of thermal and chemical equilibrium. A 
departure from equilibrium conditions may alter this conjecture. 
 
When confronting specific experimental data it is important to eliminate 
every possibility that recorded particle ratios could come from a genuine 
hadronic phase. To this end we must consider the largest possible value of 
the MIT bag constant $B$, which maximises the value of $T_0$ and, 
consequently, of the space occupied by the hadronic domain. From [21,22] 
we are informed that 
\begin{equation} 
B^{1/4}=(145-235)\;\;\;MeV\;. 
\end{equation} 
The above range of values of $B$ corresponds to a maximum critical 
temperature in the range of 
\begin{equation} 
T_0\approx (145-183)\;\;\;MeV\;. 
\end{equation} 
 
To exhaust all available space for the hadronic phase we adopt the upper 
value of $T_0$, i.e. $ T_0\approx 183\;\;MeV$. It is of interest to note that 
this particular value of $T_0$ leads to $\mu_{s\;cr}$ at the end of the 
hadronic phase (for the case $\alpha=4$) with values closest to zero, but 
without becoming negative (see Fig. 10a). A relevant study pertaining to 
data on particle ratios from the experiments at AGS (Brookhaven) and SPS 
(CERN), within the theoretical bounds set by this value of $T_0$, is 
currently in progress [23].

\newpage 
{\large{\bf 7. Concluding Remarks}} 
 
Our basic effort, in this paper, was to enlarge the SBM scheme so as to 
incorporate the strangeness quantum number. Such an attempt is {\it a 
priori} justified in view of the association of strangeness with one of the key 
experimental signals for the QGP state of matter, expected to be achieved in 
ultrarelativistic nucleus-nucleus collisions. Our efforts were directed towards 
the determination, in the 3-dimensional $(T,\mu_q,\mu_s)$ space, of : 1) 
the critical surface, determined by the bootstrap equation, which sets the 
limits of the hadronic phase and 2) the $<S>=0$ surface, relevant to 
hadronic processes, which is determined from the partition function. 
 
We restricted our attention to two different versions of the extended SBM, 
corresponding to the values $\alpha=2$ and $\alpha=4$ of the exponent 
which enters the asymptotic expression for the mass spectrum $\rho(m)$, cf. 
equ. (23). The first has dominated SBM studies in the past, in the absence of 
the strangeness quantum number. The numerical studies performed in this 
paper revealed decisive, physically motivated, advantages of the $\alpha=4$ 
version of the strangeness-containing SBM. In addition, one now has the 
phenomenological advantage of uniquely relating the MIT bag constant to 
the critical temperature $T_0$. 
 
Finally, we considered the issue of produced particle ratios for cases where 
at least one of the two species carries strangeness. The relevant study has 
been conducted within the $\alpha=4$ version of the SBM. Taking into 
account the constraint furnished by $<S>=0$, we have presented numerical 
results pertaining to particle ratios along a fixed direction in the $(T,\mu_q)$ 
plane which show: (a) for points well inside the hadronic domain the ratios 
coincide with those of the ``ideal hadron gas model'', (b) as the critical 
surface is approached (bootstrap input) departure from the ideal description 
begin to show up. We also presented results on critical particle ratios, as a 
function of the (critical) temperature, along the intersection curve of the $<S>=0$ 
and critical surfaces. 
 
Clearly, the present effort constitutes only a first step towards an 
increasingly complete investigation of multi-hadron production systems, 
coming from heavy-ion collisions, under the assumption of thermalization 
on the hadronic side. Obvious theoretical refinements are: (1) The 
differentiation between the up-down quantum numbers (isospin breaking 
effects) which will introduce separate chemical potentials $\mu_u$, $\mu_d$ 
in place of $\mu_q$. On the hadronic side this entails the extension 
$\lambda_B\rightarrow(\lambda_B,\lambda_Q)$, where $\lambda_Q$ is the 
fugacity corresponding to total charge and (2) The introduction into our 
description of the variable 
$\gamma_s^2=\lambda_s^{neq}\lambda_{\bar{s}}^{neq}\neq1$, where ``$neq$''
stand for ``non-equilibrium'', pertaining to a measure 
of {\it partial} strangeness equilibrium. 
 
On a broader basis, of course, lies the problem of how to relate information 
obtained from the hadronic side on the one hand, with QCD-related input 
coming from the QGP phase of matter on the other. The outlook for 
investigations of this nature clearly has a longer range prospective. 
 
As already mentioned, work is now in progress which addresses itself to 
actual experimental data on (strange) particle ratios. The aim of such an 
effort is to identify, always according to the $\alpha=4$ version of the 
extended SBM, whether the multiparticle source, which furnishes a 
particular set of data, lies inside or outside the bounds of the phase of 
hadrons. In a forthcoming paper [23] we shall present the results of such 
analyses.

\vspace{2cm} 
\begin{center} 
{\large{\bf Acknowledgements}} 
\end{center} 
 
{\it We wish to thank Prof. R. Hagedorn for useful discussions. A. S. K. also 
wishes to thank Dr. C. G. Papadopoulos for discussion and guidance in 
certain parts of the numerical calculations. 
 
This work was partially supported by the programme $\rm \Pi ENE\Delta$ 
No. 1361.1674/31-1-95, Gen. Secretariat for Research and Technology, 
Hellas.}

\vspace{2cm} 
\begin{center} 
{\large{\bf Figure Captions}} 
\end{center} 
\newtheorem{f}{Figure} 
\begin{f} 
\rm $G$ as a function of $\varphi$. The branch point sets a limit on the 
hadronic 
description of matter. 
\end{f} 
\begin{f} 
\rm (a) Intersections of planes of constant s-quark chemical potential $\mu_s$
with the critical surface $\varphi(T,\mu_q,\mu_s)=\ln 4-1$ for of $\alpha=2$
and for $T_0=180$ MeV. (b) Intersections of planes of constant q-quark 
chemical potential $\mu_q$ with the critical surface 
$\varphi(T,\mu_q,\mu_s)=\ln 4-1$ for $\alpha=2$ and for $T_0=180$ MeV. 
\end{f} 
\begin{f}
\rm (a) Variation of the intersection of the plane of constant s-quark 
chemical potential $\mu_s=0$ with the critical surface 
$\varphi(T,\mu_q,\mu_s)=\ln 4-1$, for $\alpha=2$, for different values of 
$T_0$. (b) Variation of the intersection of the plane of constant q-quark 
chemical potential $\mu_q=0$ with the critical surface 
$\varphi(T,\mu_q,\mu_s)=\ln 4-1$, for $\alpha=2$, for different values of 
$T_0$.
\end{f}
\begin{f}
\rm Projection on the plane $(T,\mu_s)$ of intersections of planes of 
constant q-quark fugacity $\lambda_q$ ($\mu_q/T=0.4$) with the surface 
$<S>=0$ for different values of $T_0$, for the $\alpha=2$ version of the 
SBM.
\end{f}
\begin{f}
\rm (a) Projection on the plane $(\mu_q,\mu_s)$ of the intersection of the 
critical surface and the surface $<S>=0$ for the value of $\alpha=2$. (b) 
Projection on the plane $(T,\mu_s)$ of the intersection of the critical surface 
and the surface $<S>=0$ for the $\alpha=2$ version of the SBM.
\end{f}
\begin{f}
\rm (a) Intersections of planes of constant s-quark chemical potential 
$\mu_s$ with the critical surface $\varphi(T,\mu_q,\mu_s)=\ln 4-1$ for 
$\alpha=4$ and for $T_0=180$ MeV. (b) Intersections of planes of constant 
q-quark chemical potential $\mu_q$ with the critical surface 
$\varphi(T,\mu_q,\mu_s)=\ln 4-1$ for $\alpha=4$ and for $T_0=180$ MeV.
\end{f}
\begin{f}
\rm (a) Variation of the intersection of the plane of constant s-quark 
chemical potential $\mu_s=0$ with the critical surface 
$\varphi(T,\mu_q,\mu_s)=\ln 4-1$ for $\alpha=4$, for different values of 
$T_0$. (b) Variation of the intersection of the plane of constant q-quark 
chemical potential $\mu_q=0$ with the critical surface 
$\varphi(T,\mu_q,\mu_s)=\ln 4-1$, for $\alpha=4$, for different values of 
$T_0$.
\end{f}
\begin{f}
\rm (a) Comparison of the intersections of planes of constant s-quark 
chemical potential $\mu_s$ with the critical surface 
$\varphi(T,\mu_q,\mu_s)=\ln 4-1$ for the cases $\alpha=2$ and $\alpha=4$ 
and for $T_0=180$ MeV. (b) Comparison of the intersections of planes of 
constant q-quark chemical potential $\mu_q$ with the critical surface 
$\varphi(T,\mu_q,\mu_s)=\ln 4-1$ for the cases $\alpha=2$ and $\alpha=4$ 
and for $T_0=180$ MeV.
\end{f}
\begin{f}
\rm Projection on the plane $(T,\mu_s)$ of intersections of planes of 
constant q-quark fugacity $\lambda_q$ ($\mu_q/T=0.4$) with the surface 
$<S>=0$ for different values of $T_0$, for $\alpha=4$.
\end{f}
\begin{f}
\rm (a) Projection on the plane $(\mu_q,\mu_s)$ of the intersection of the 
critical surface and the surface $<S>=0$ for $\alpha=4$. (b) Projection on 
the plane $(T,\mu_s)$ of the intersection of the critical surface and the 
surface $<S>=0$ for $\alpha=4$.
\end{f}
\begin{f}
\rm A three-dimensional perspective of our results for the $\alpha=4$ 
version of the SBM. The set of rising and subsequently falling curves 
correspond to intersections of planes of constant $\lambda_q$ with the 
surface $<S>=0$. The intersection of this surface with the critical one, for 
$T_0=183$ MeV, is the contour which lies on the edges of the former 
curves.
\end{f}
\begin{f}
\rm (a) The ratio $K^+/\pi^+$ as a function of temperature, for 
$\mu_q/T=0.4$. (b) The ratio $K^-/\pi^-$ as a function of temperature, for 
$\mu_q/T=0.4$. (c) The ratio $K^+/K^-$ as a function of temperature, for 
$\mu_q/T=0.4$. (d) The ratio $K_s^0/\pi^0$ as a function of temperature, 
for $\mu_q/T=0.4$. (e) The ratio $\Xi^+/\bar{\Lambda}$ as a function of 
temperature, for $\mu_q/T=0.4$. (f) The ratio $\Xi^-/\Lambda$ as a 
function of temperature, for $\mu_q/T=0.4$. (g) The ratio 
$\bar{\Lambda}/\Lambda$ as a function of temperature, for $\mu_q/T=0.4$. 
(h) The ratio $K_s^0/\Lambda$ as a function of temperature, for 
$\mu_q/T=0.4$. (i) The ratio $\bar{\Xi}^0/\Xi^0$ as a function of 
temperature, for $\mu_q/T=0.4$. These ratios pertain to the $\alpha=4$ 
version of the SBM.
\end{f}
\begin{f}
\rm (a) The critical value of the ratio $K^+/\pi^+$ as a function of the 
critical temperature. (b) The critical value of the ratio $K^-/\pi^-$ as a 
function of the critical temperature. (c) The critical value of the ratio 
$K^+/K^-$ as a function of the critical temperature. (d) The critical value of 
the ratio $K_s^0/\pi^0$ as a function of the critical temperature. (e) The 
critical value of the ratio $\Xi^+/\bar{\Lambda}$ as a function of the critical 
temperature. (f) The critical value of the ratio $\Xi^-/\Lambda$ as a function 
of the critical temperature. (g) The critical value of the ratio 
$\bar{\Lambda}/\Lambda$ as a function of the critical temperature. (h) The 
critical value of the ratio $K_s^0/\Lambda$ as a function of the critical 
temperature. (i) The critical value of the ratio $\bar{\Xi}^0/\Xi^0$ as a 
function of the critical temperature. These ratios pertain to the $\alpha=4$ 
version of the SBM.
\end{f}
\begin{f}
\rm (a) A situation where two experimentally measured particle ratios can 
have hadronic origin. (b) A situation where two experimentally measured 
particle ratios would imply, under the assumption of thermalisation, that 
their origin is outside the domain of hadrons.
\end{f}

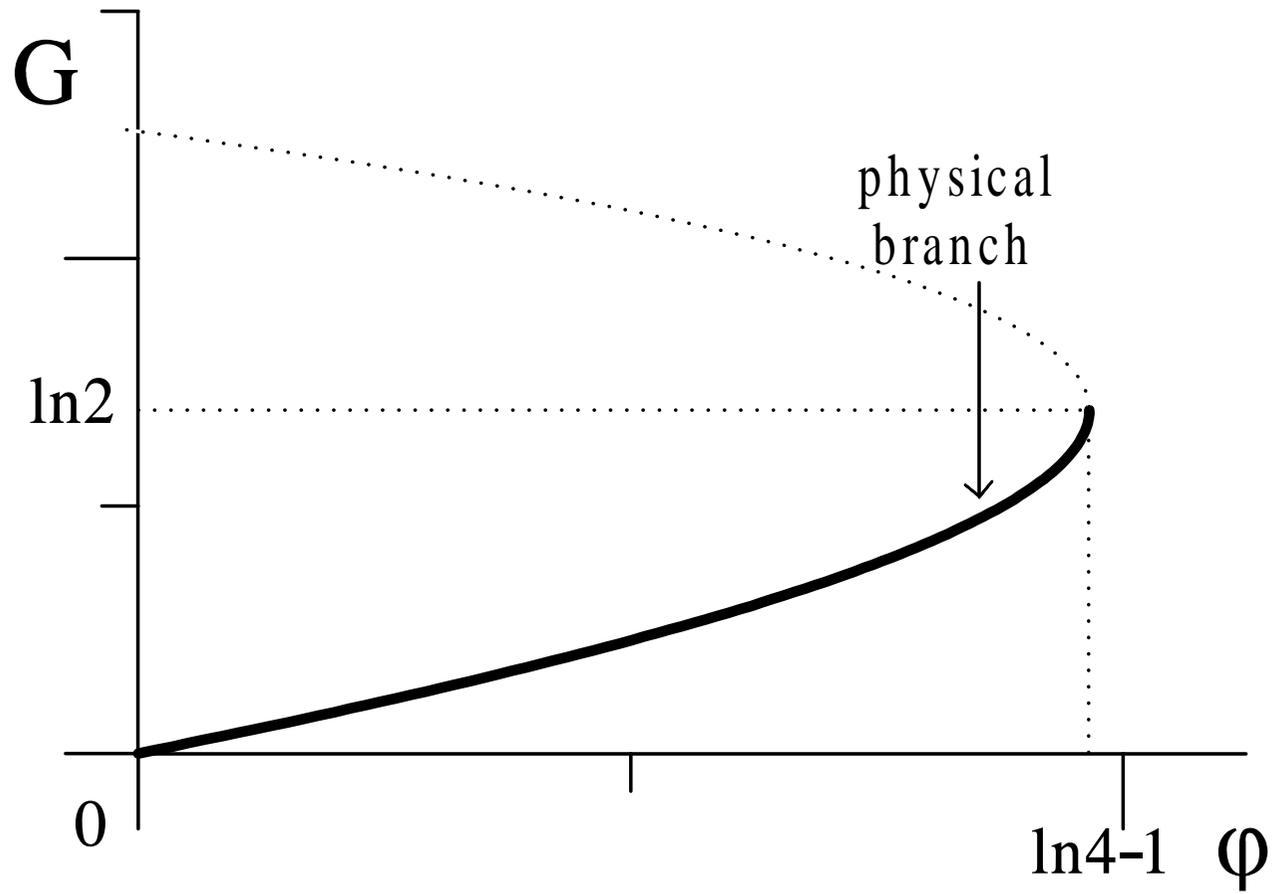
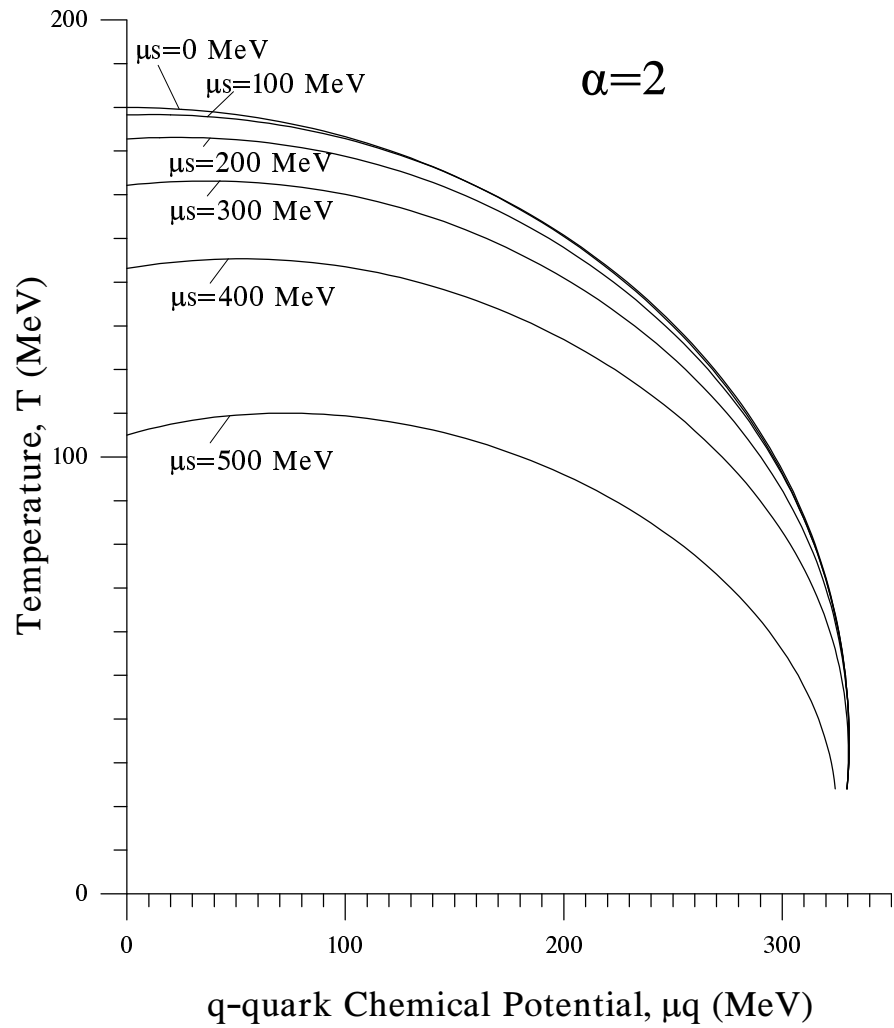
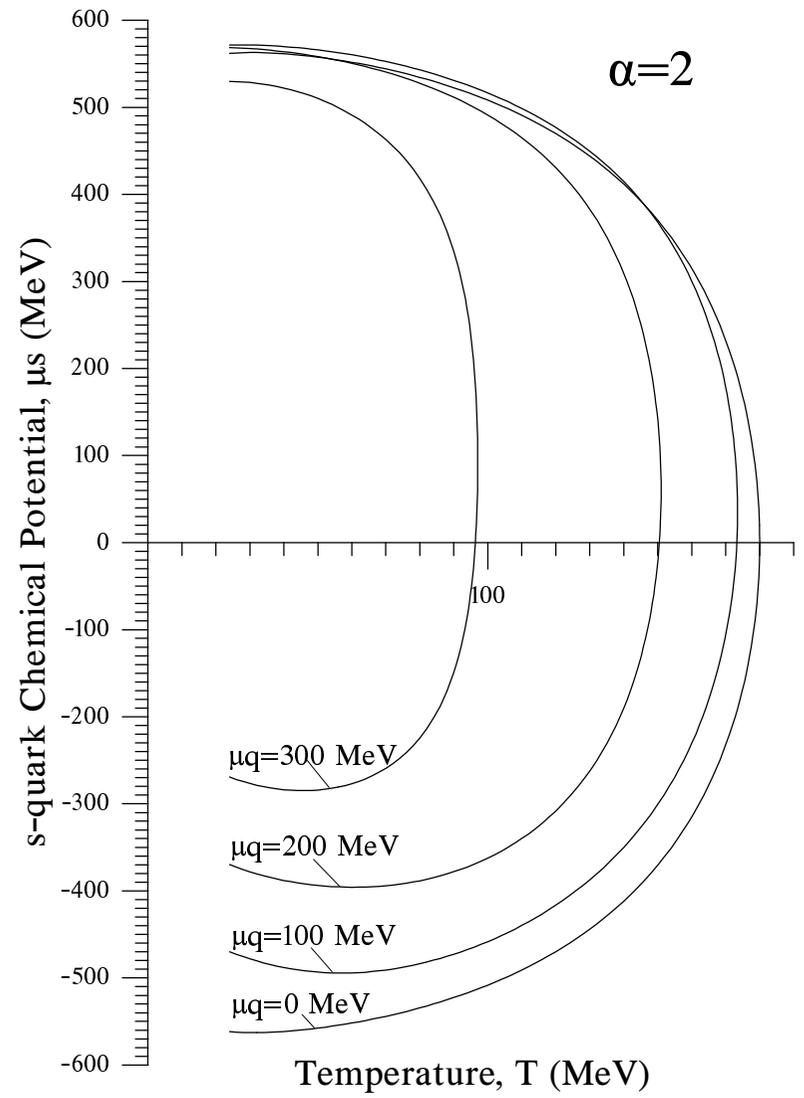
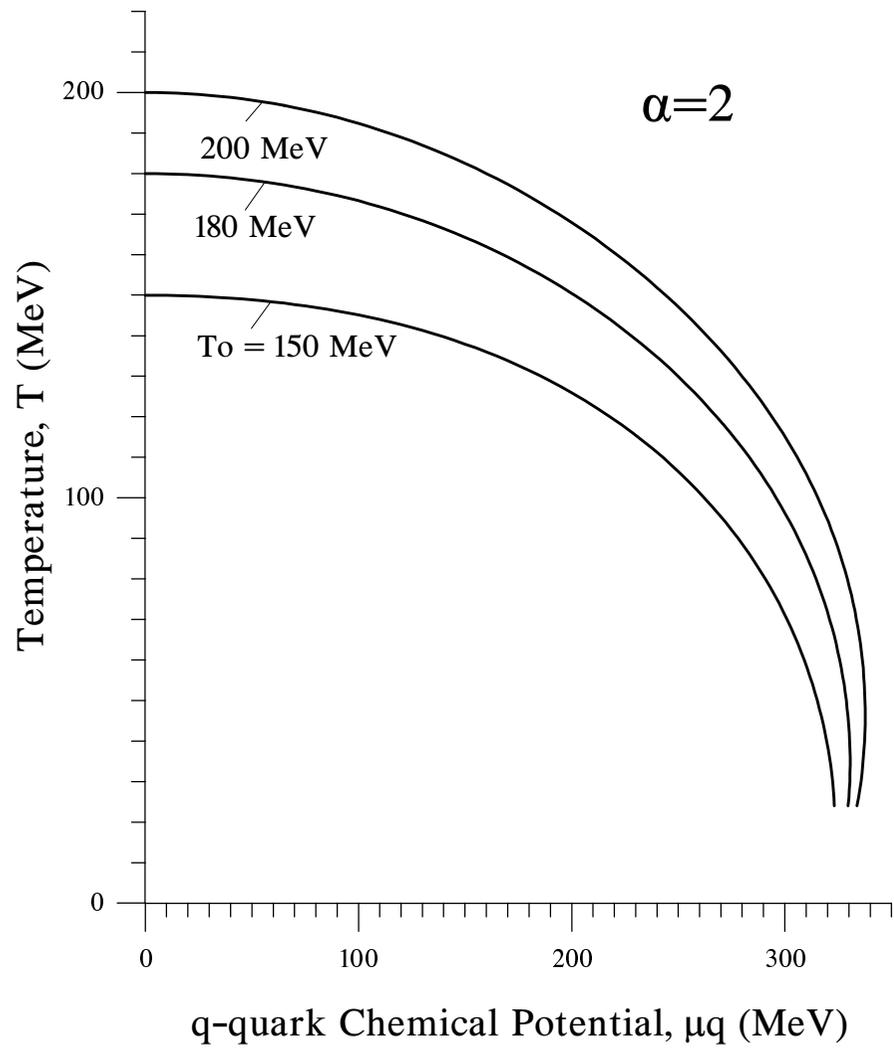
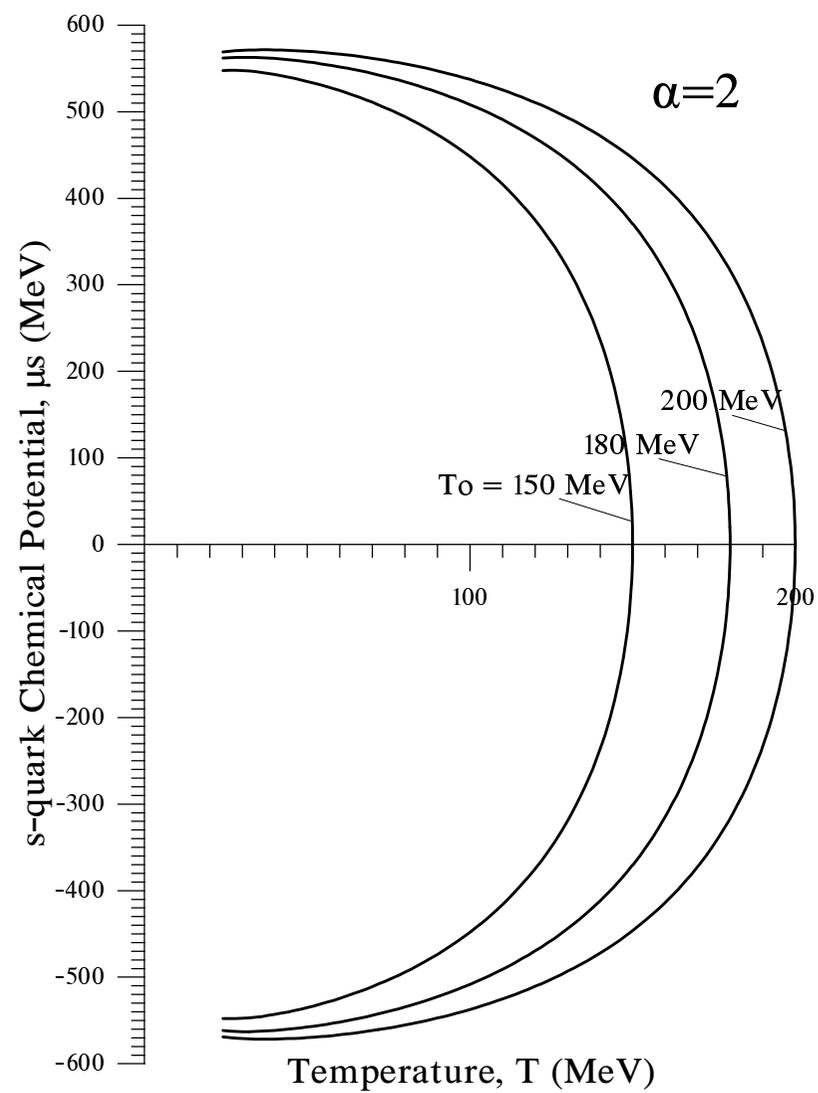
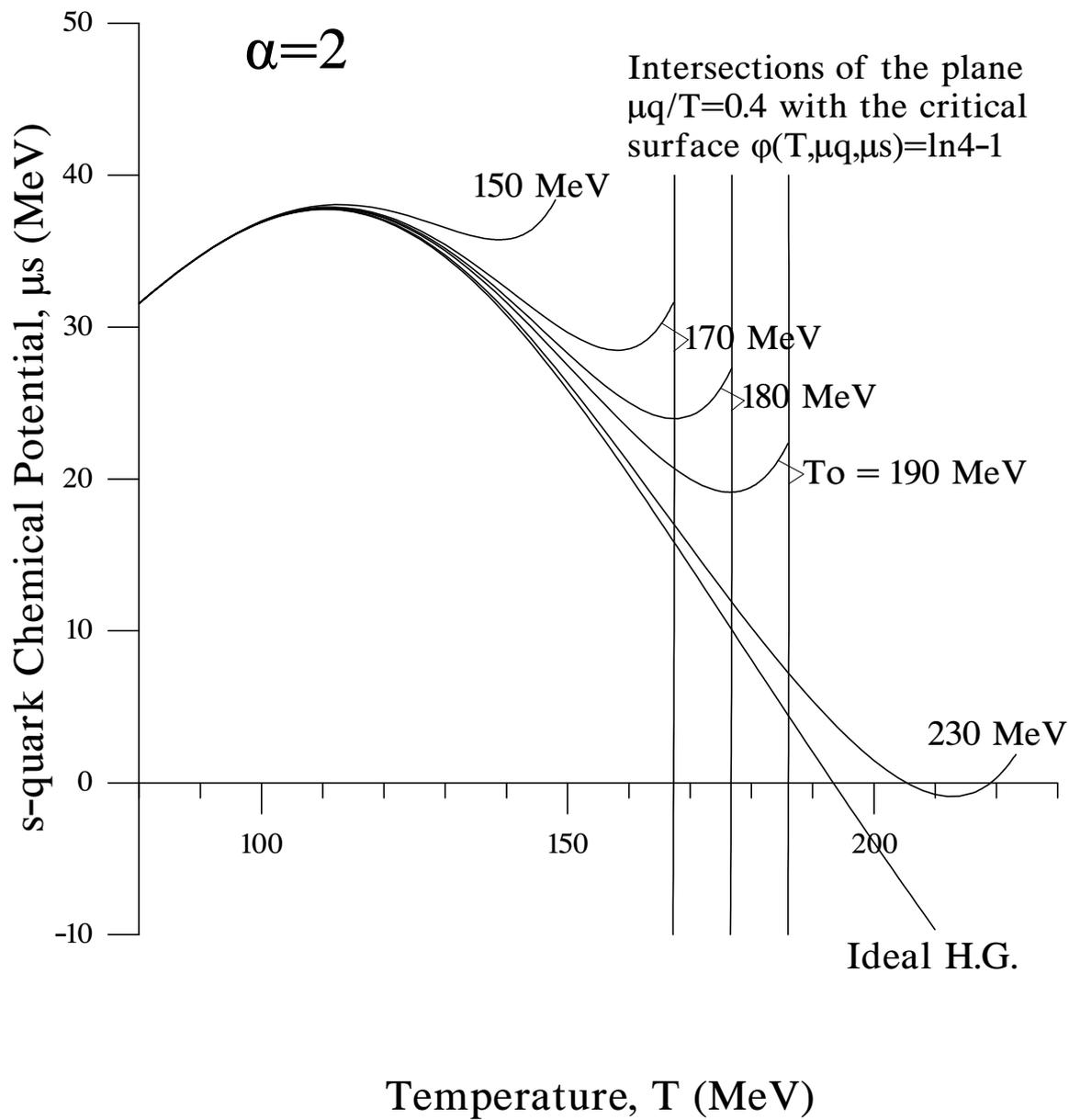
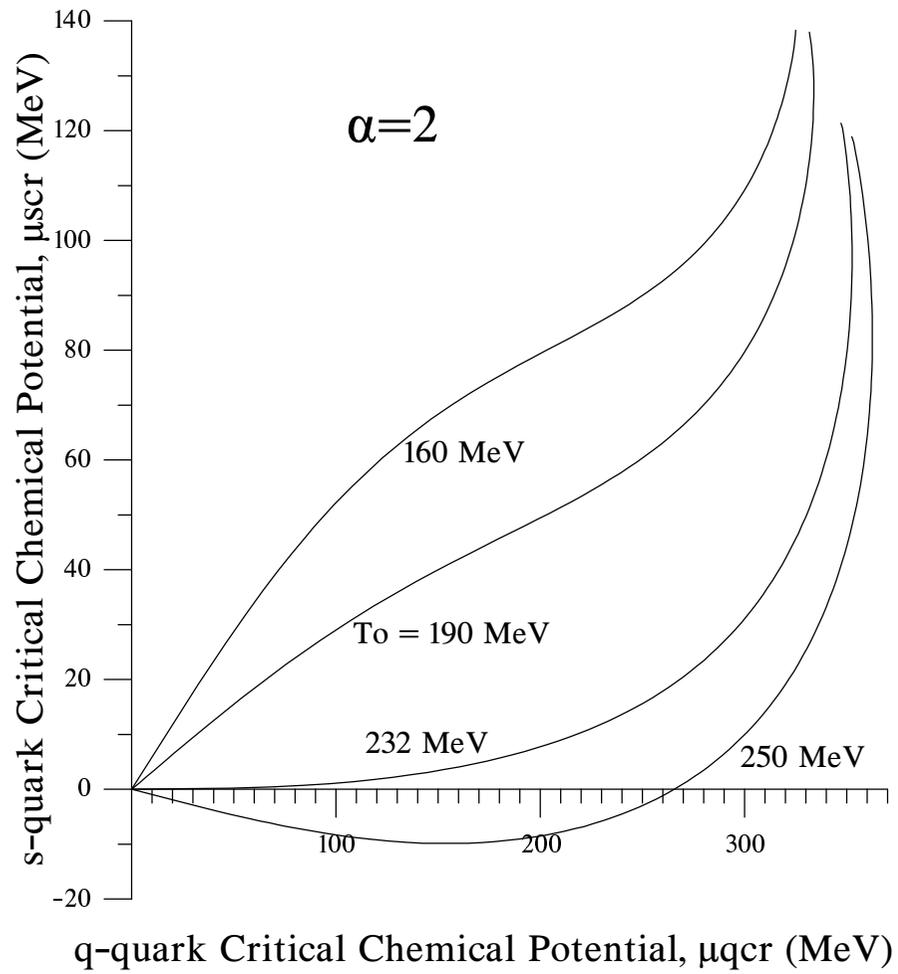
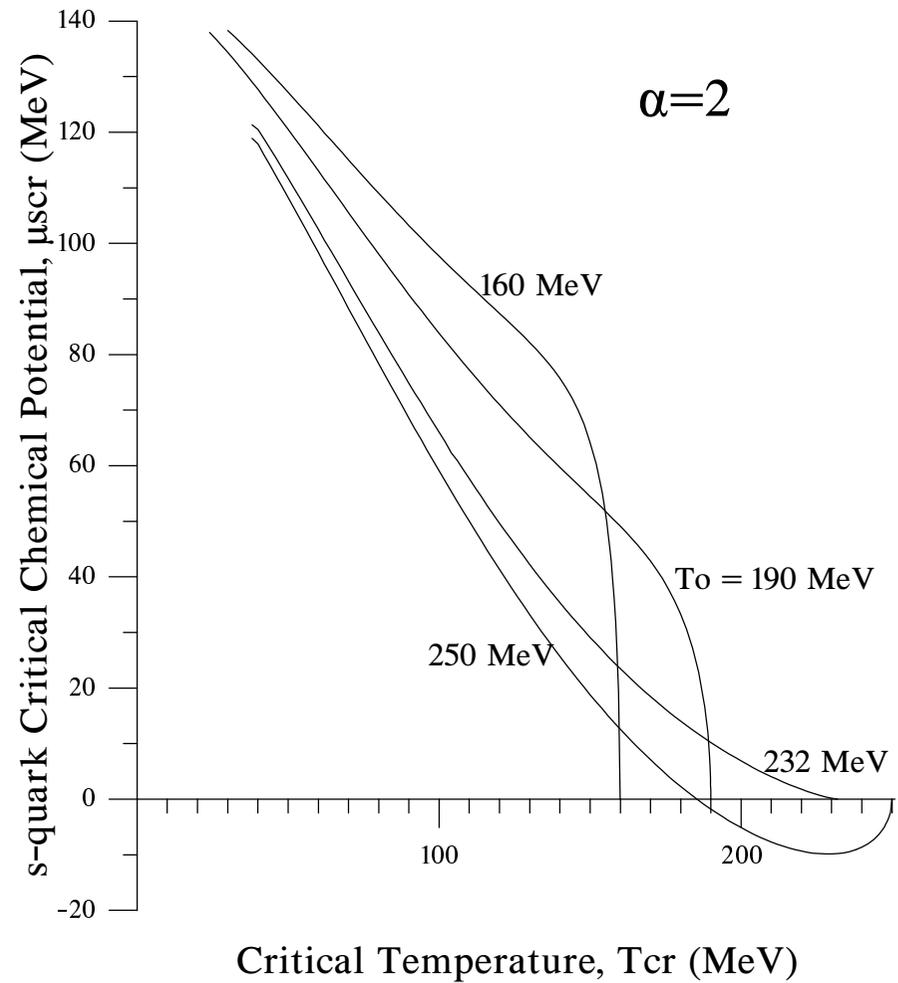
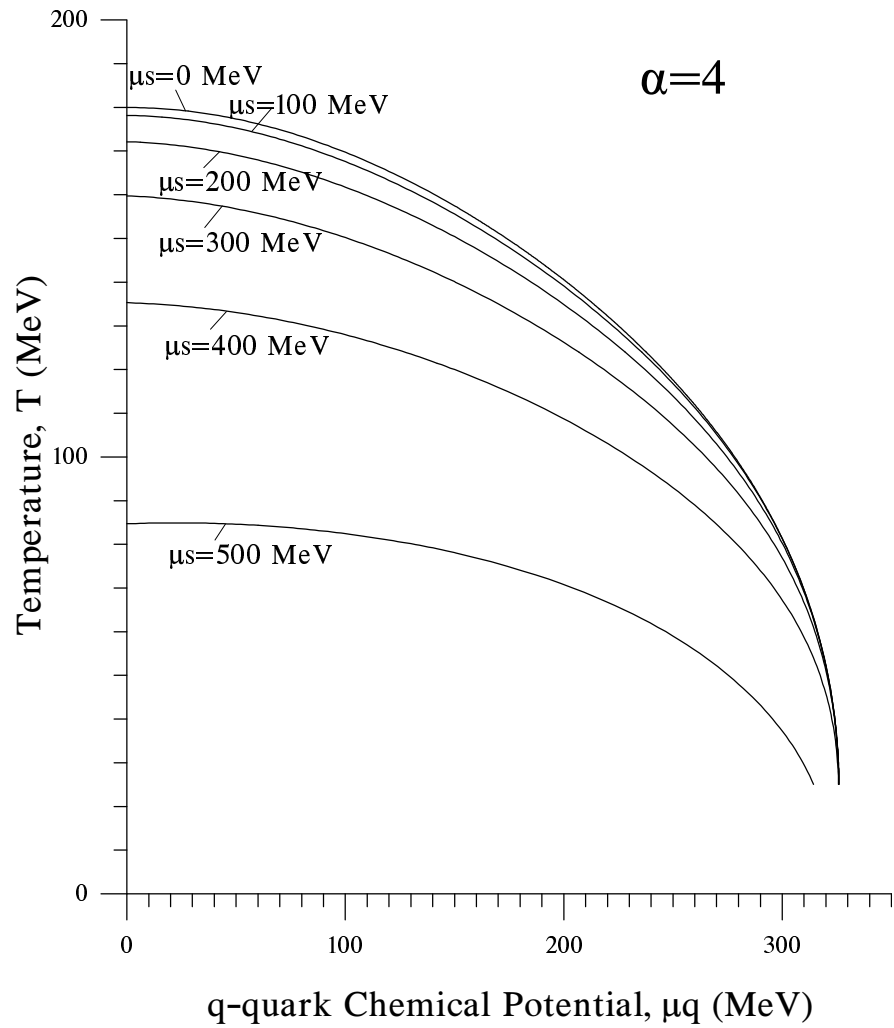
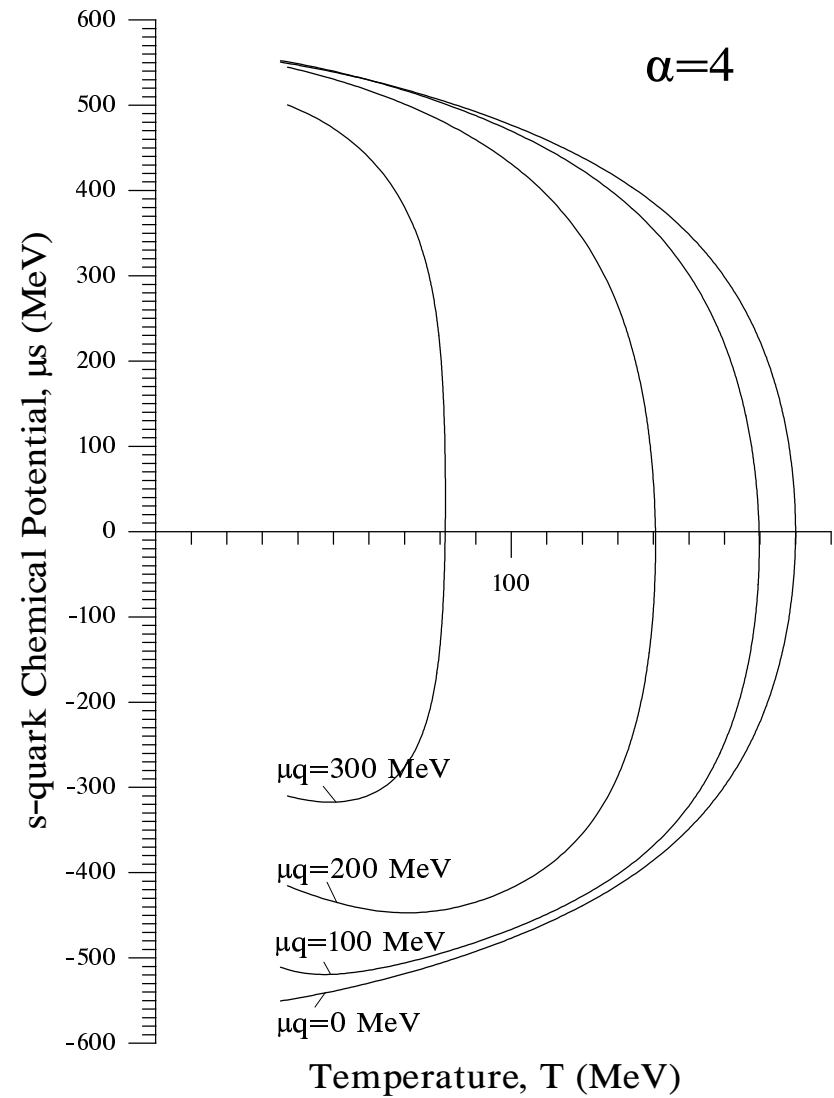
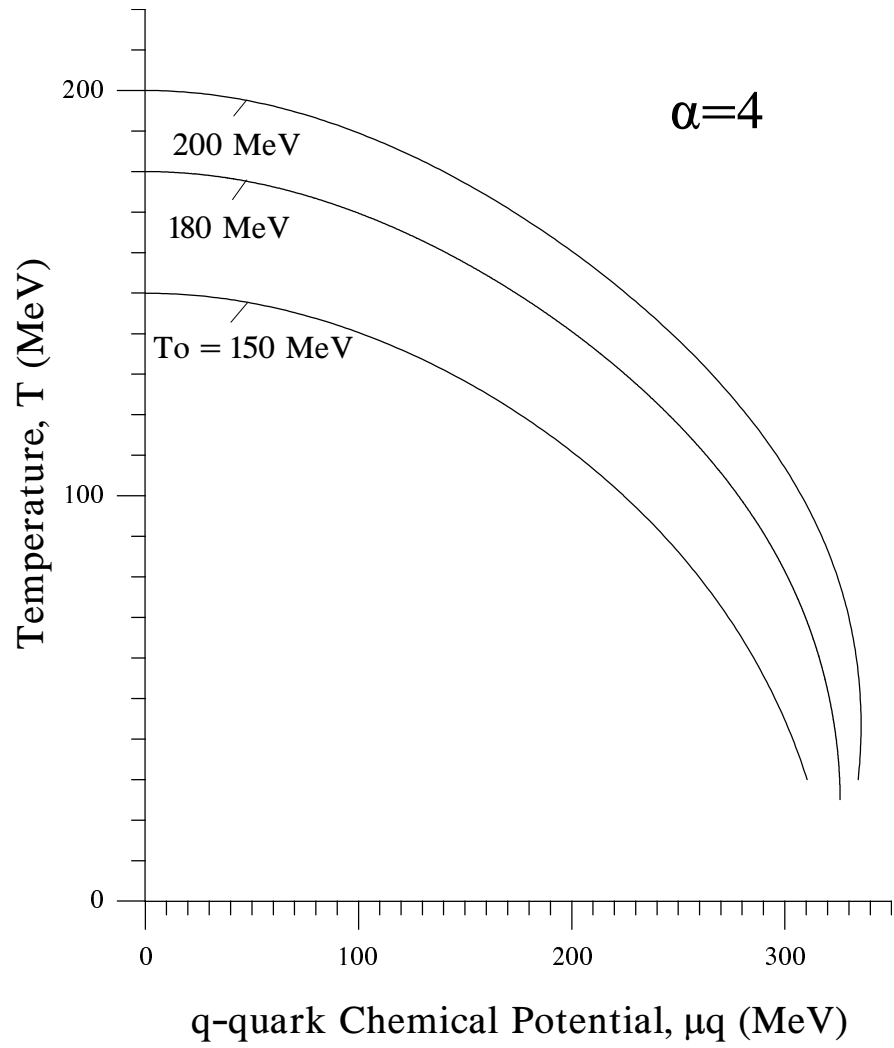
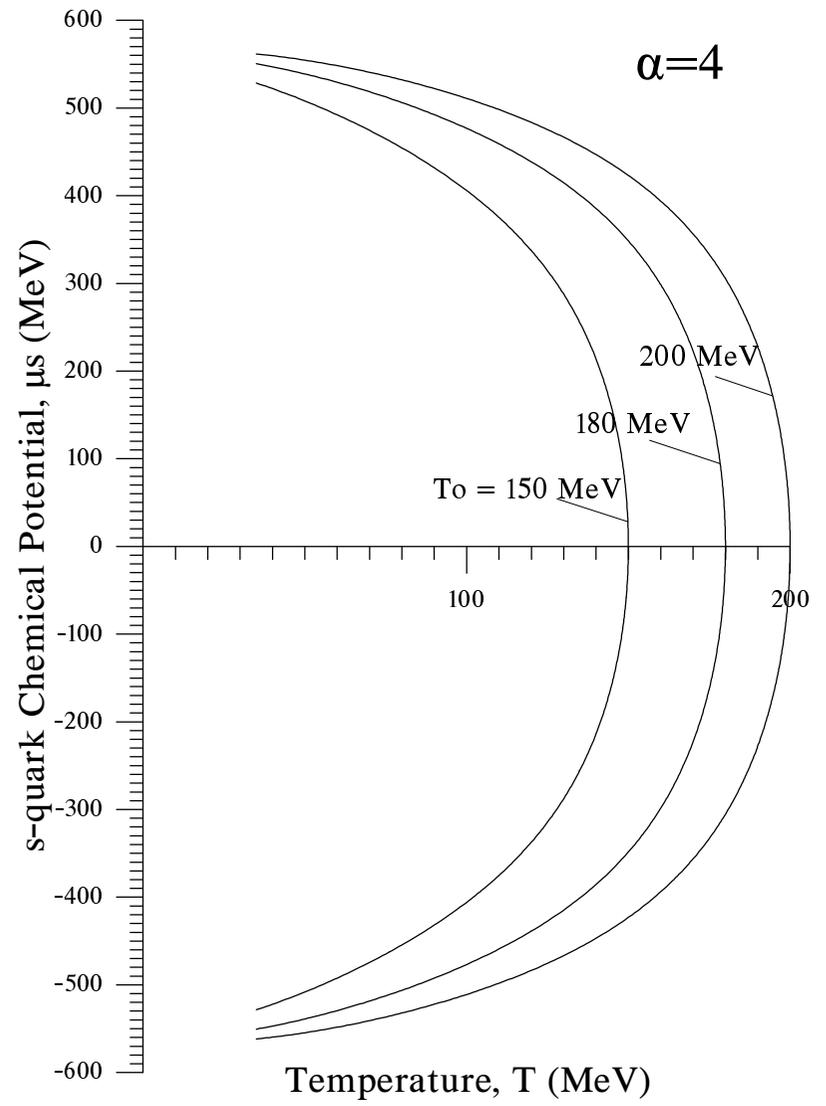
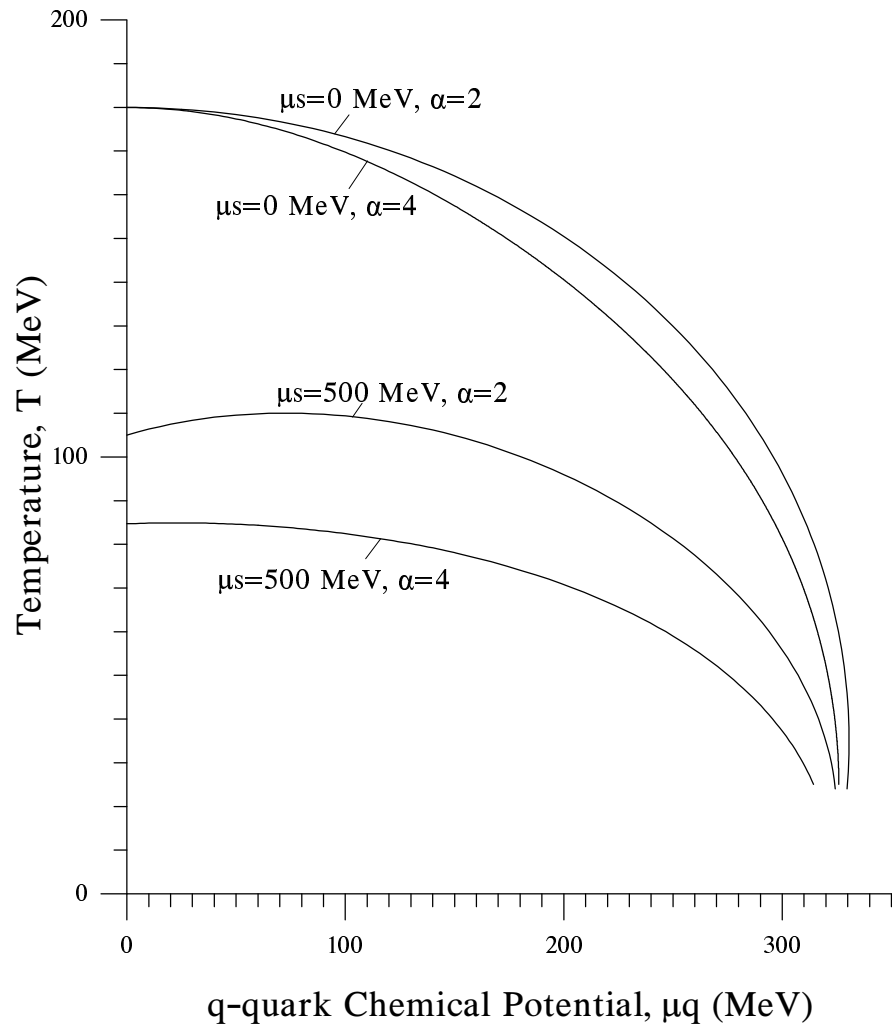
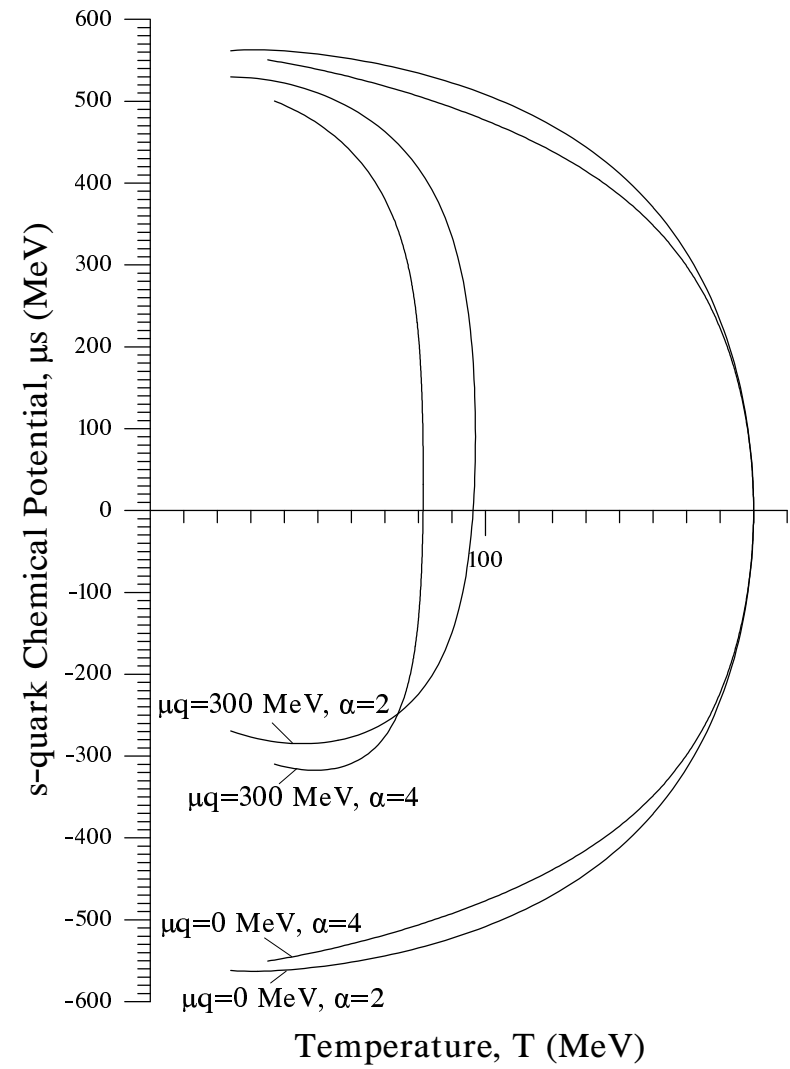
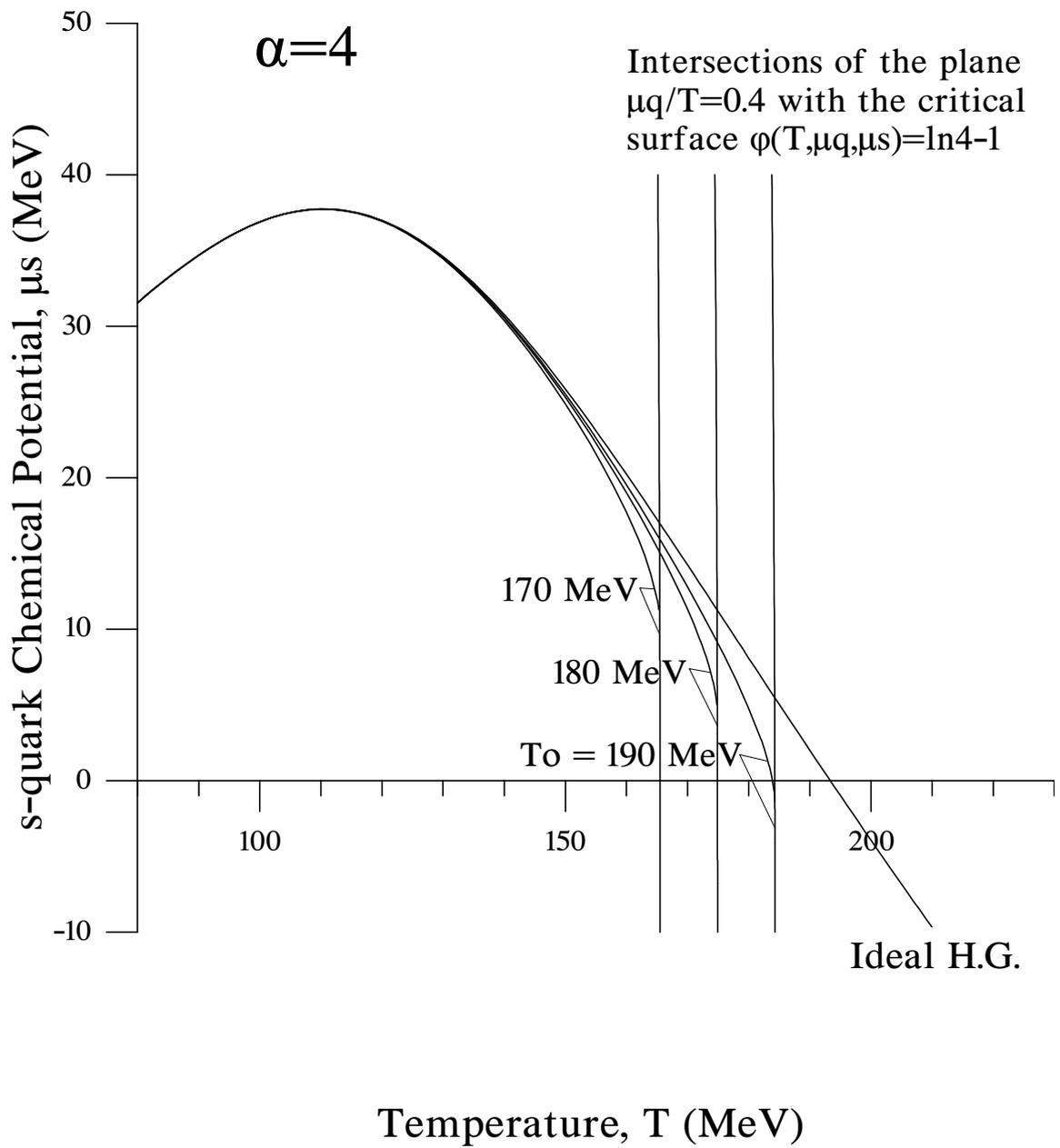
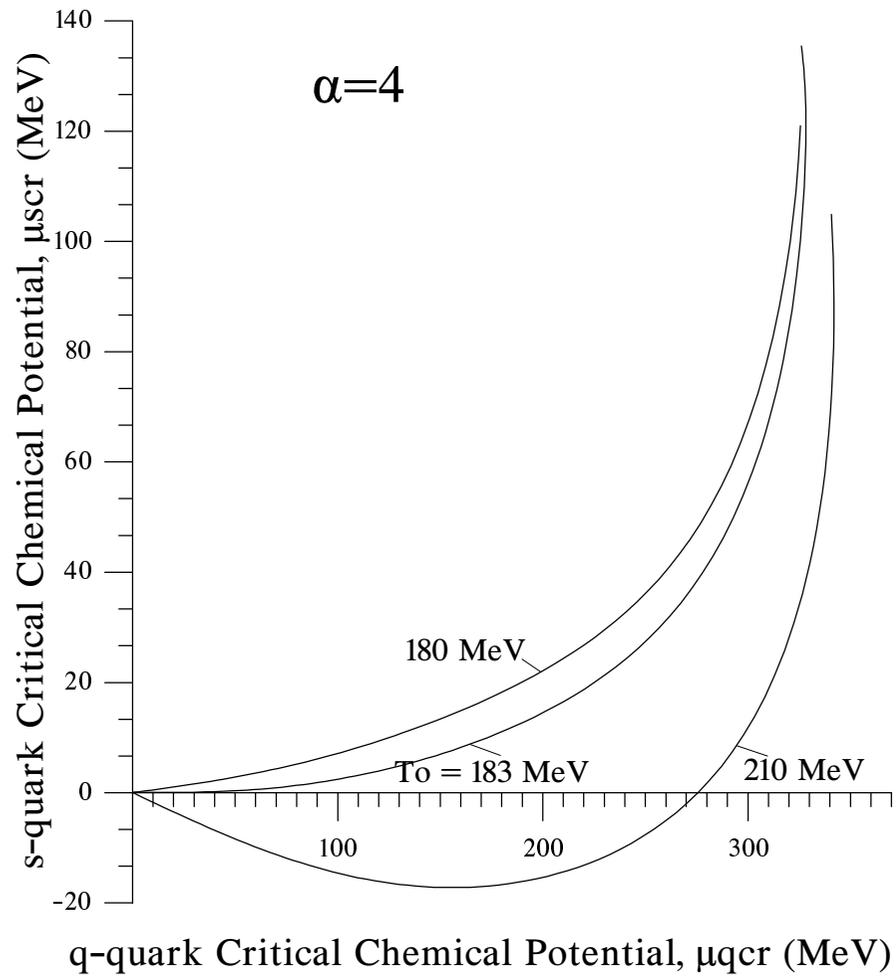
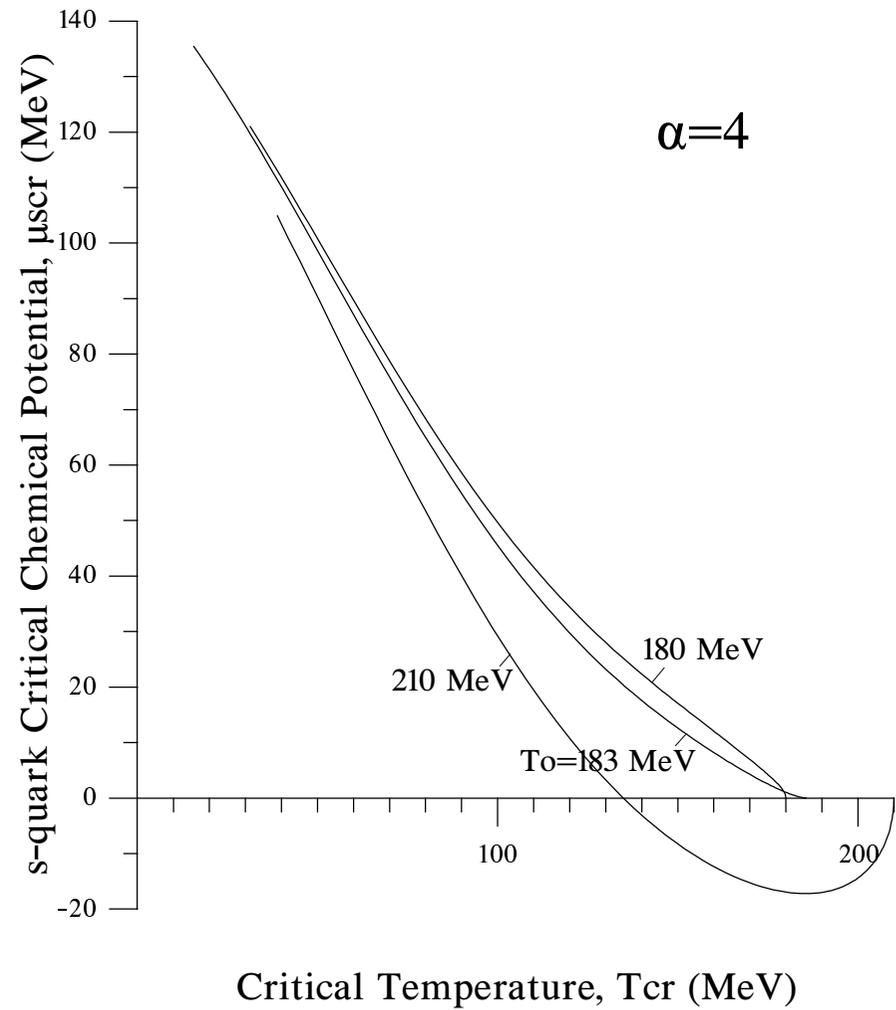
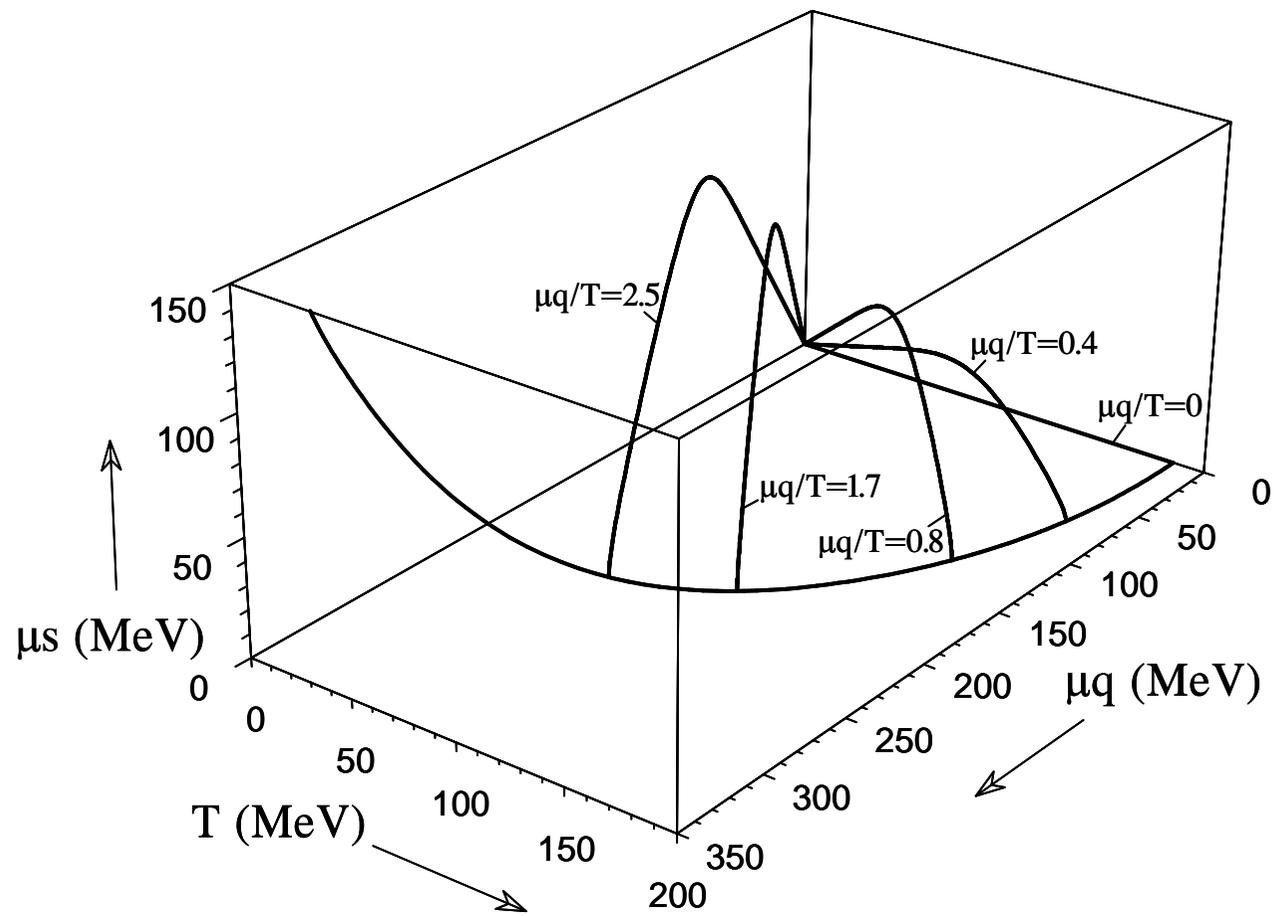
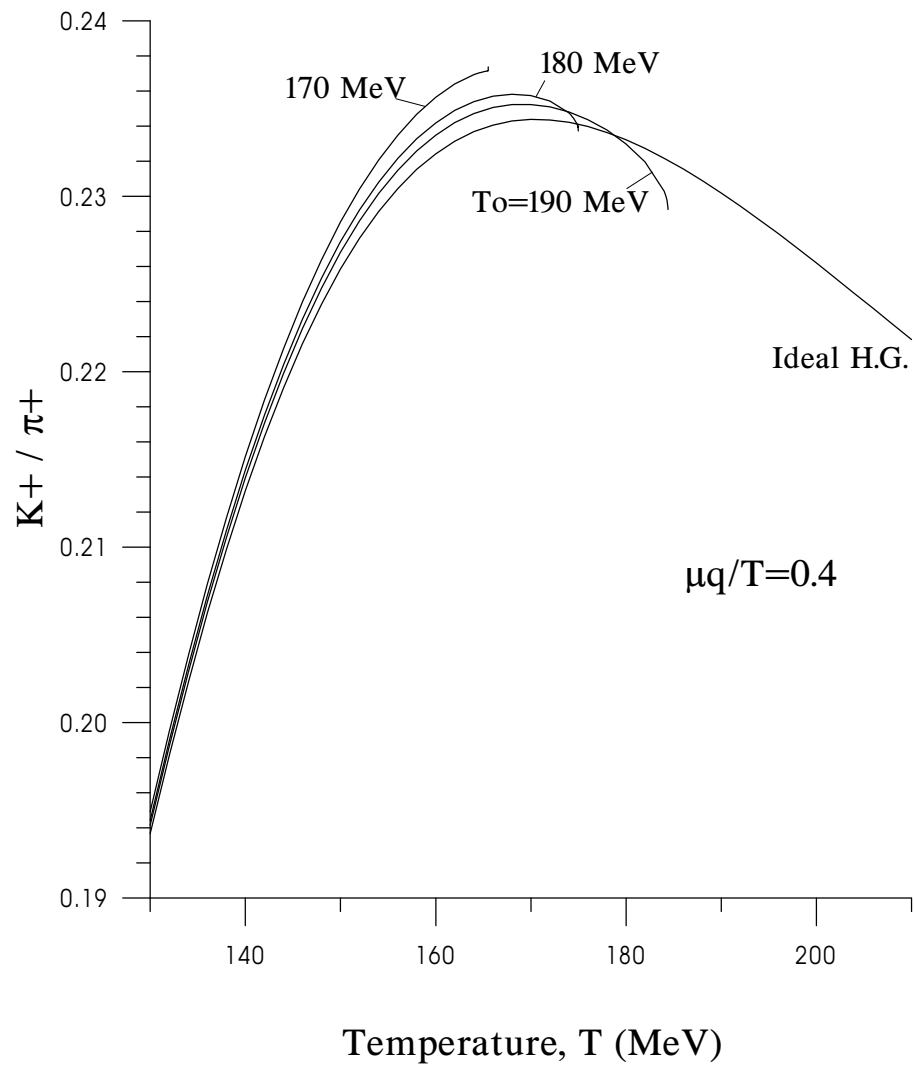
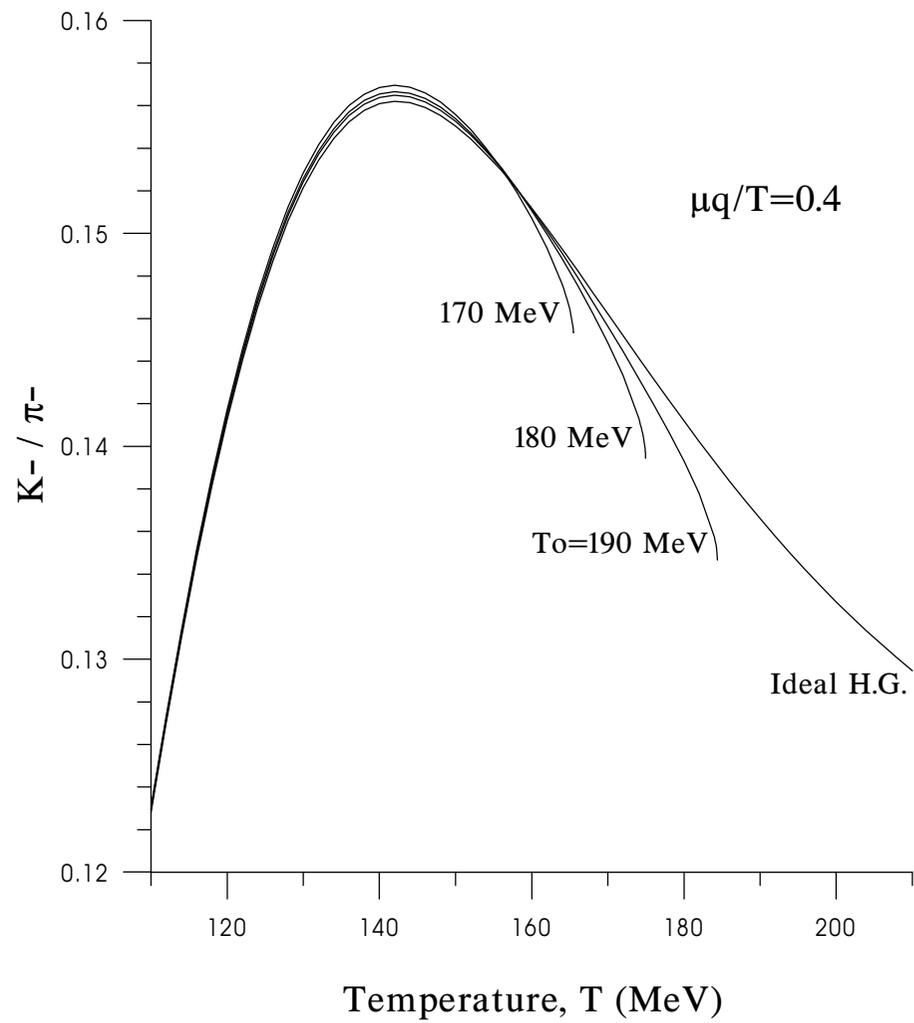
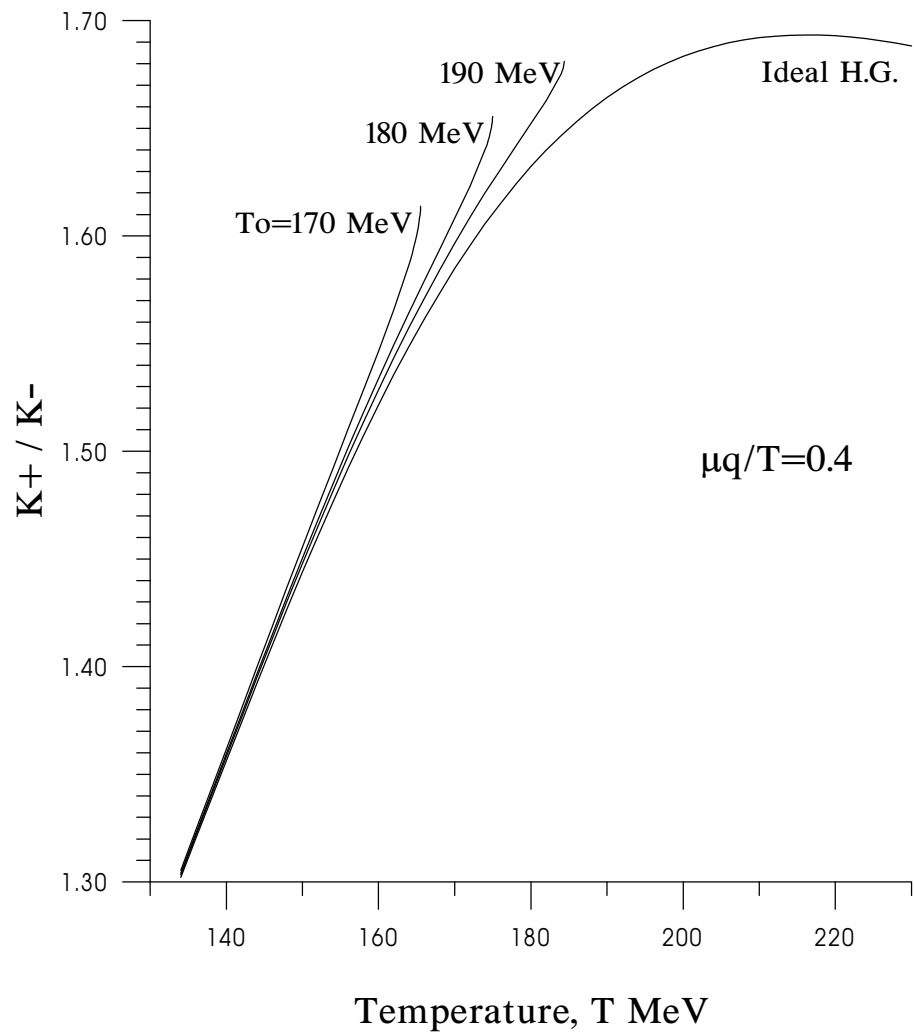
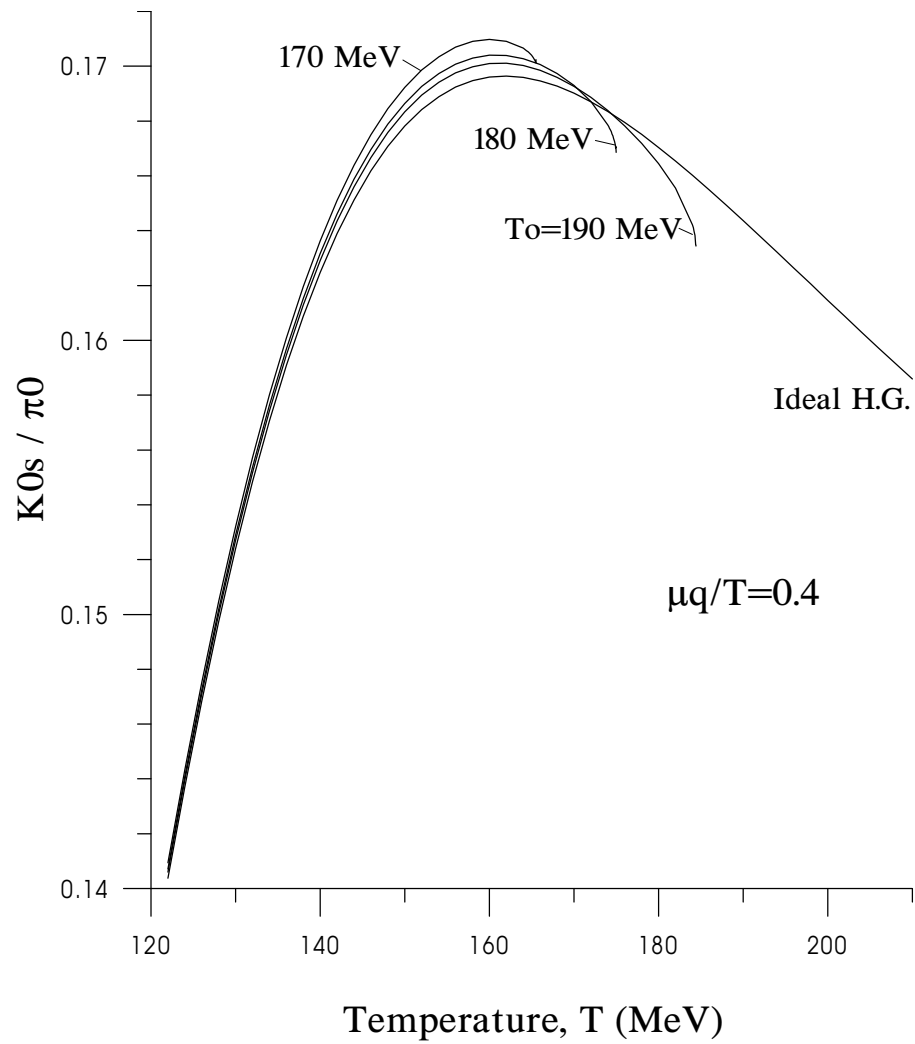
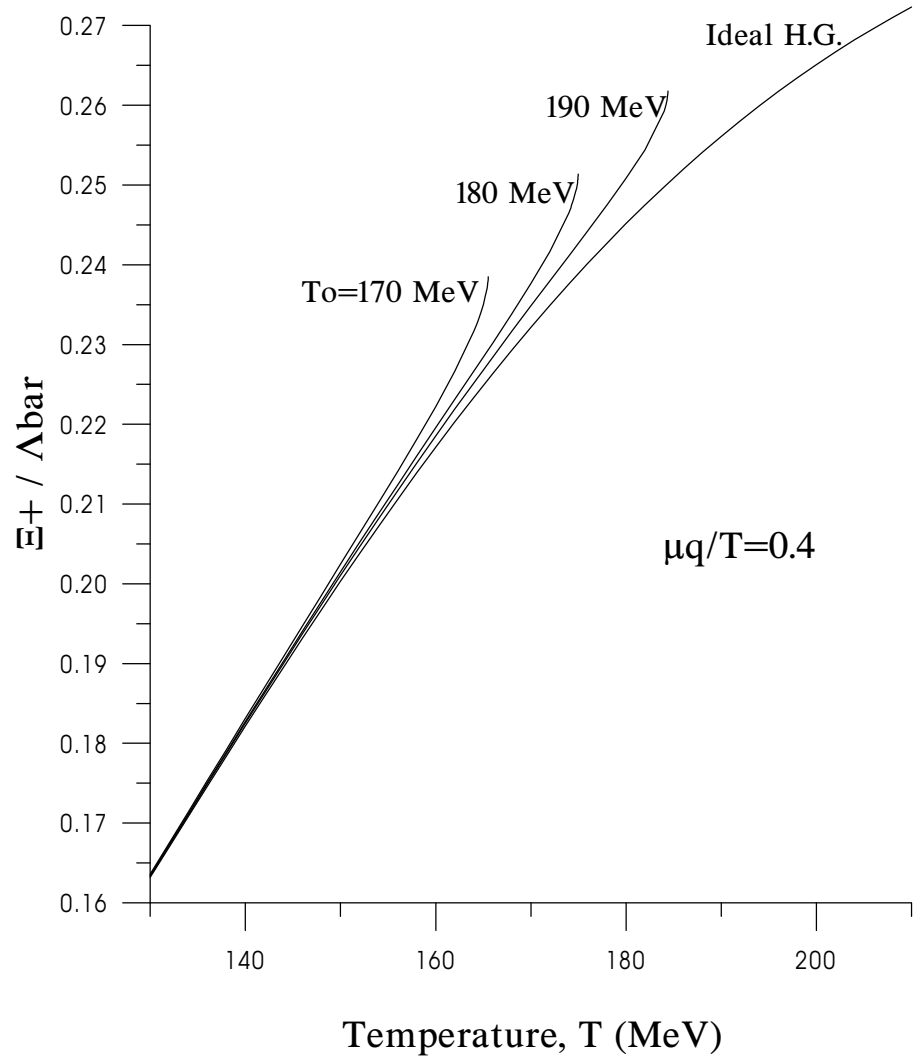
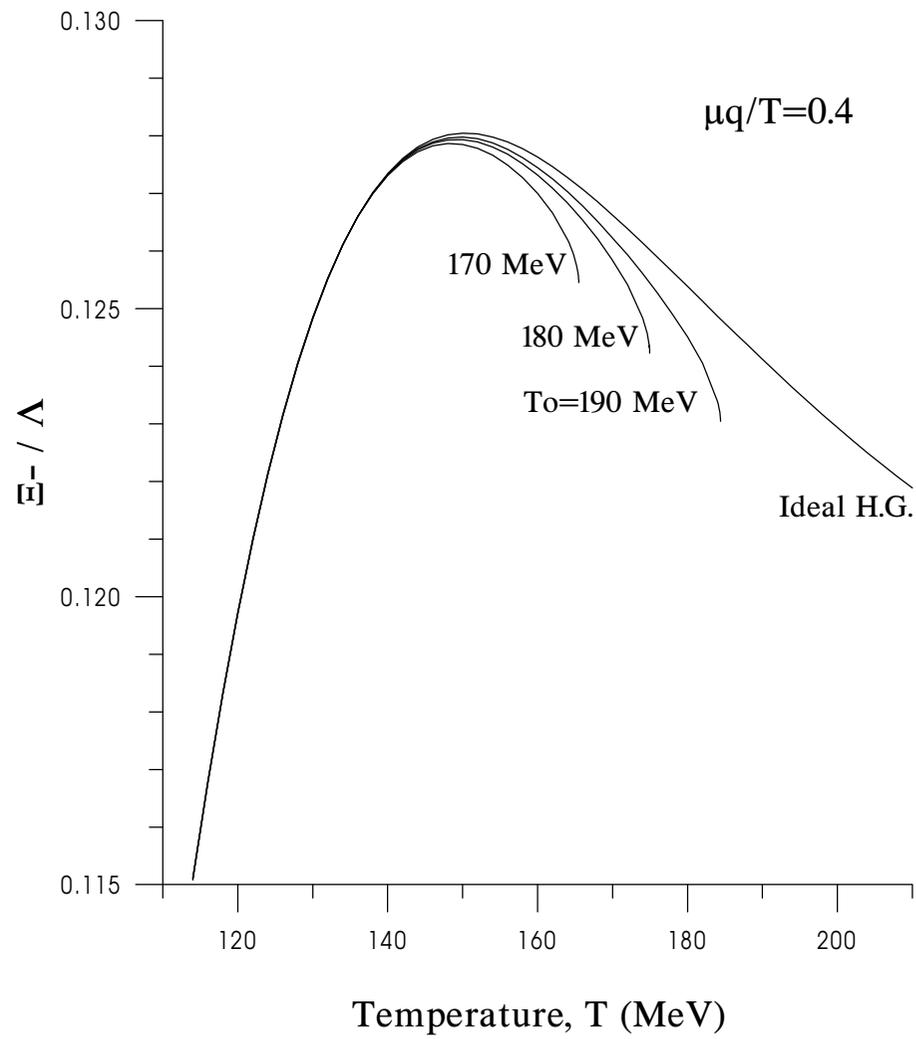
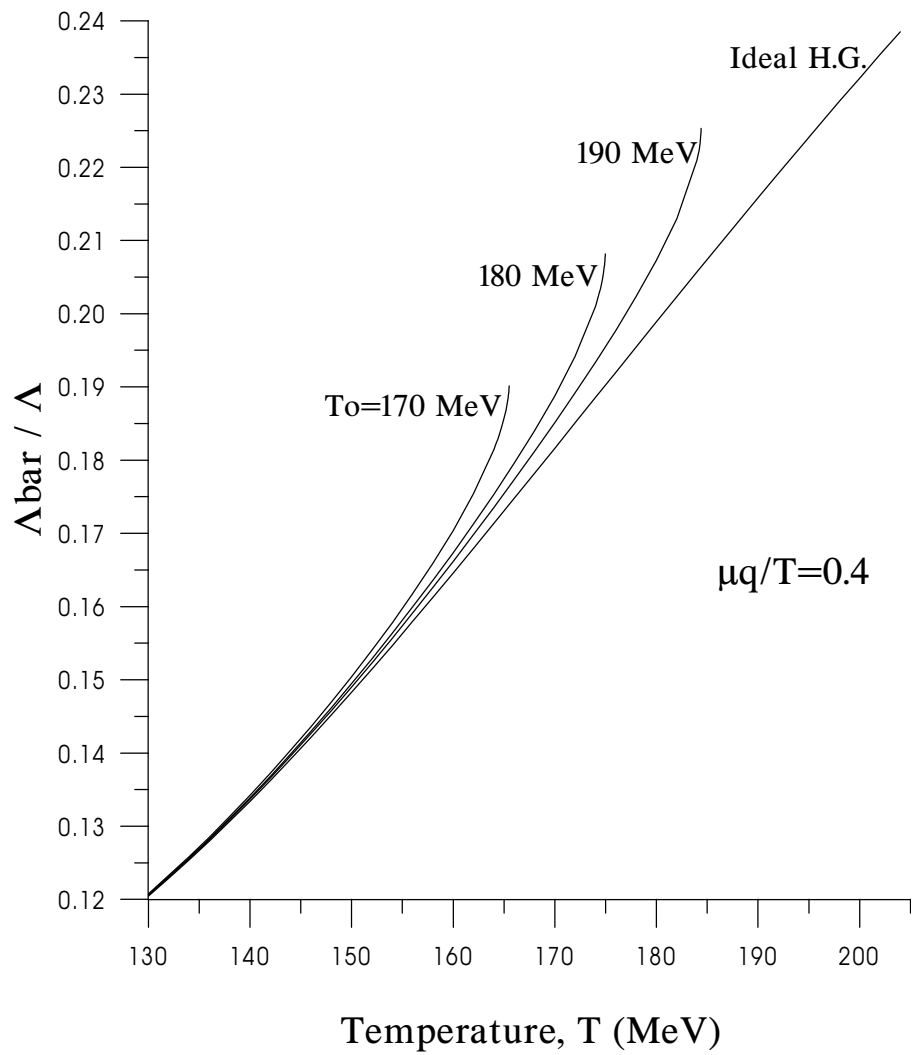
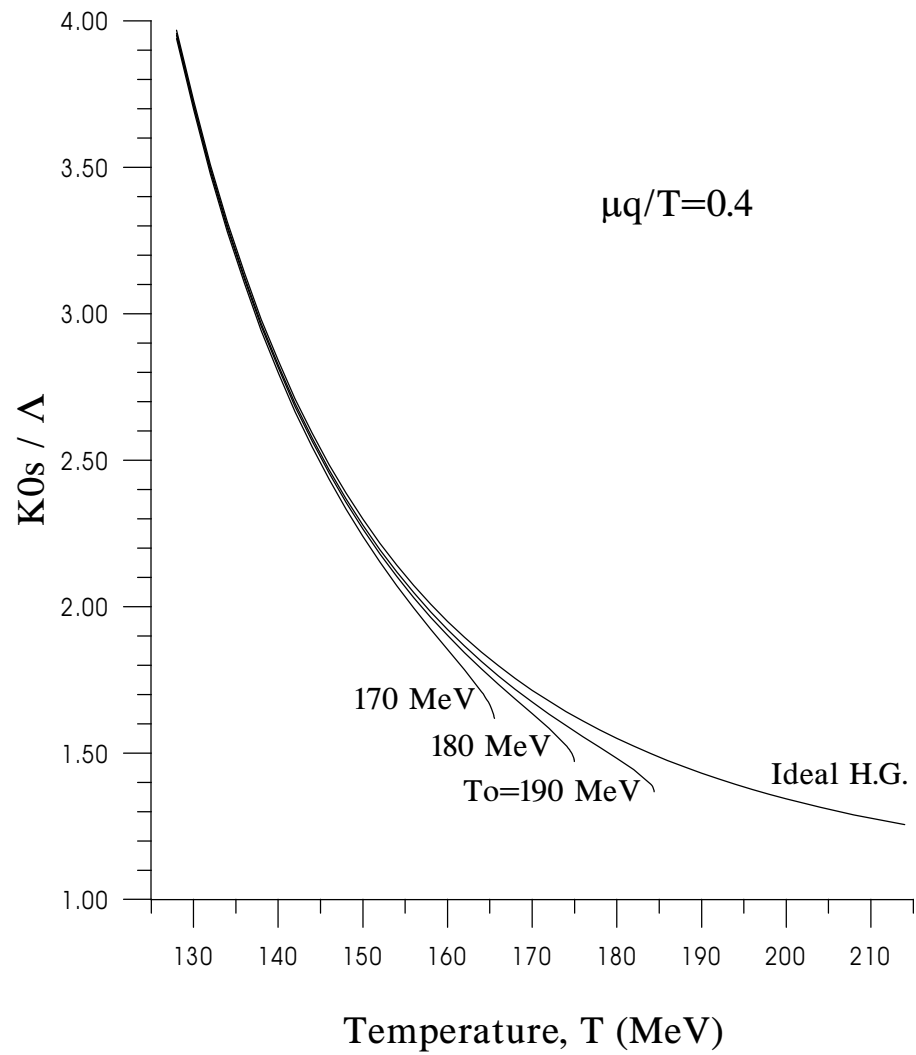
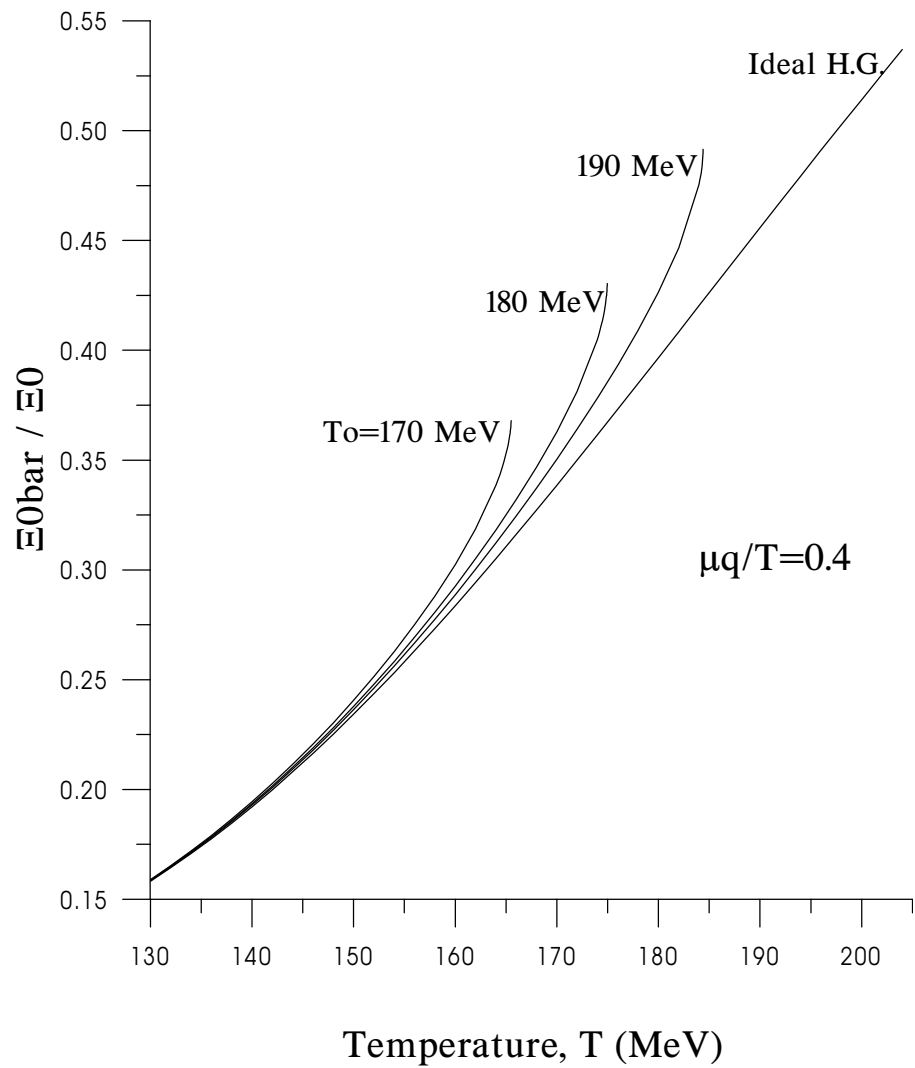
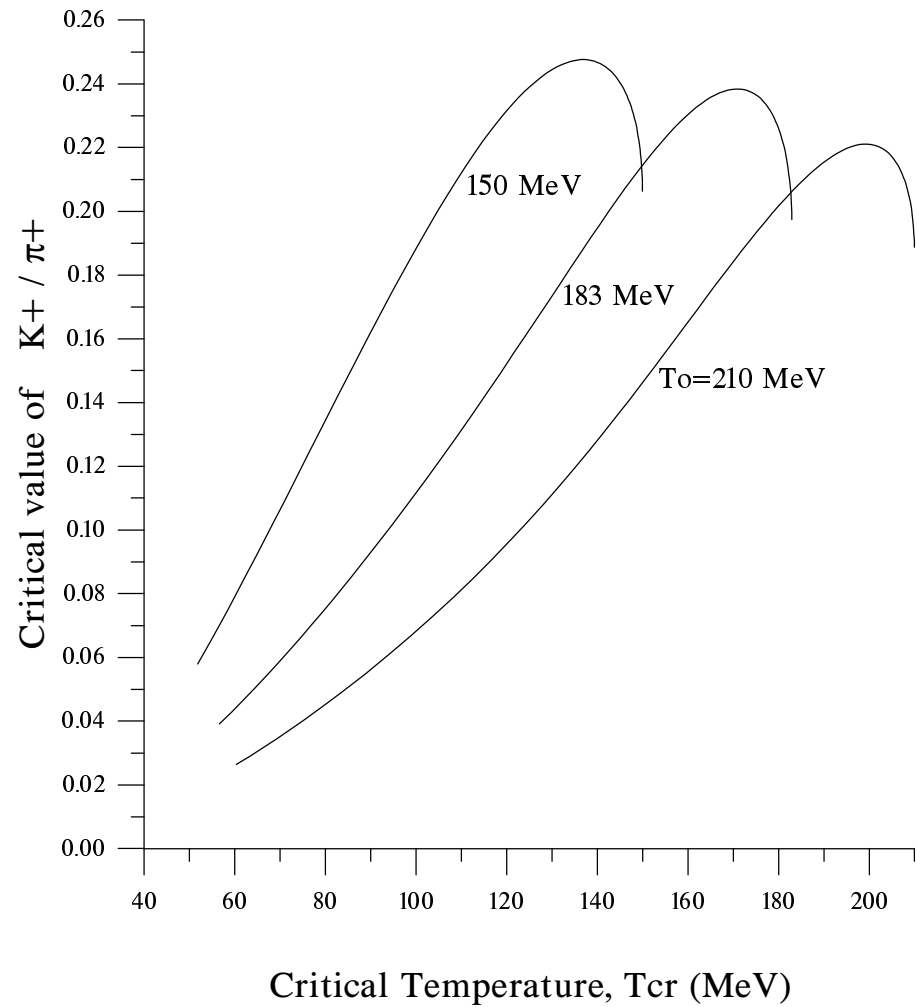
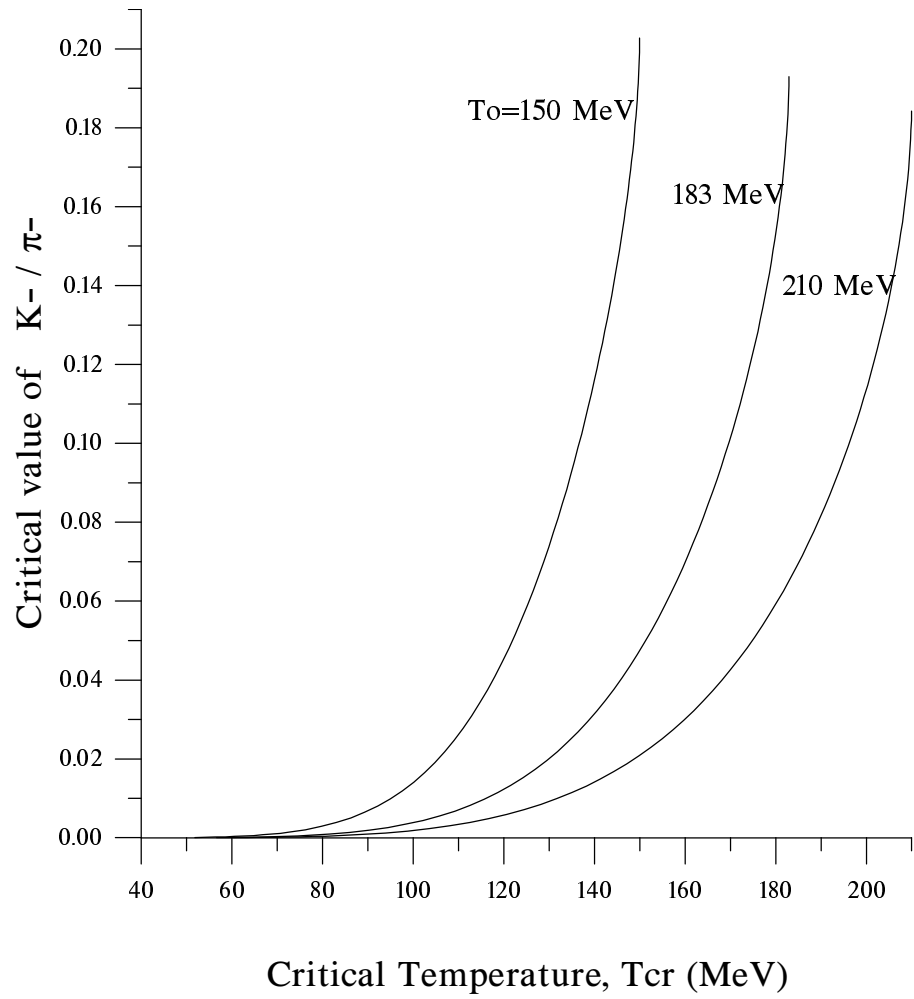
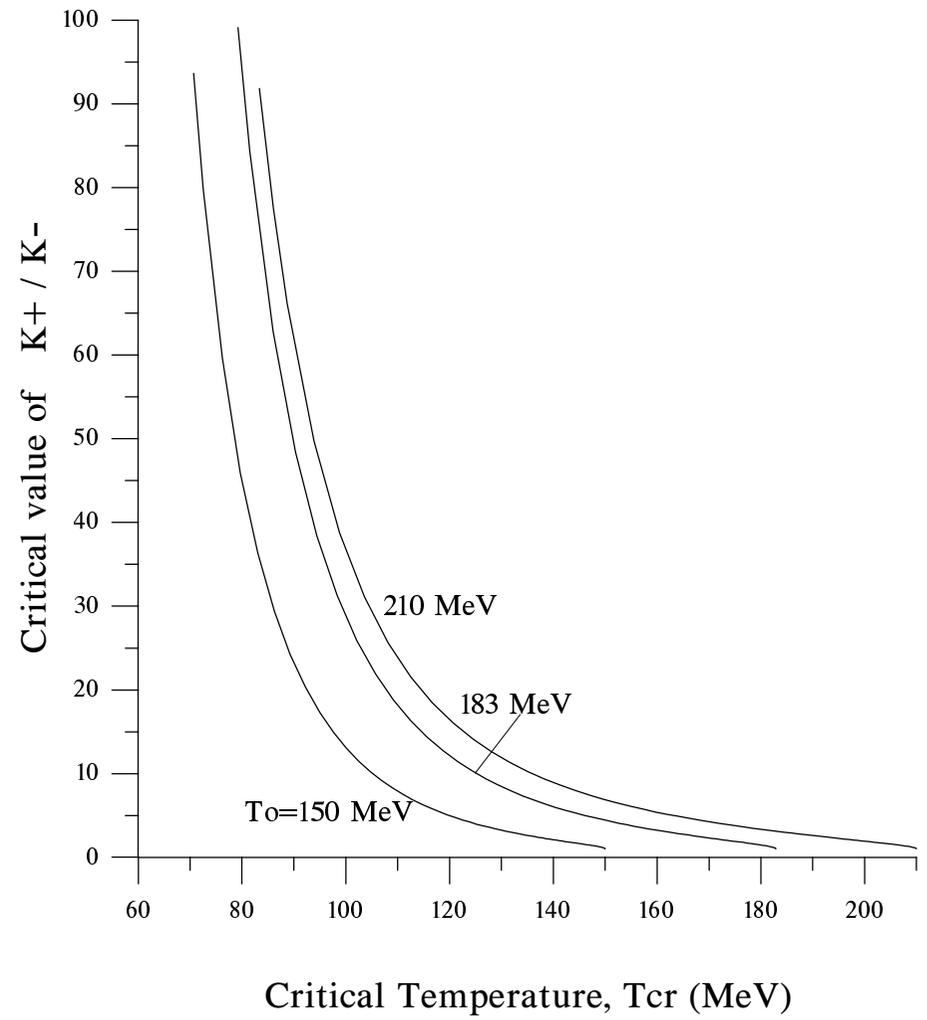
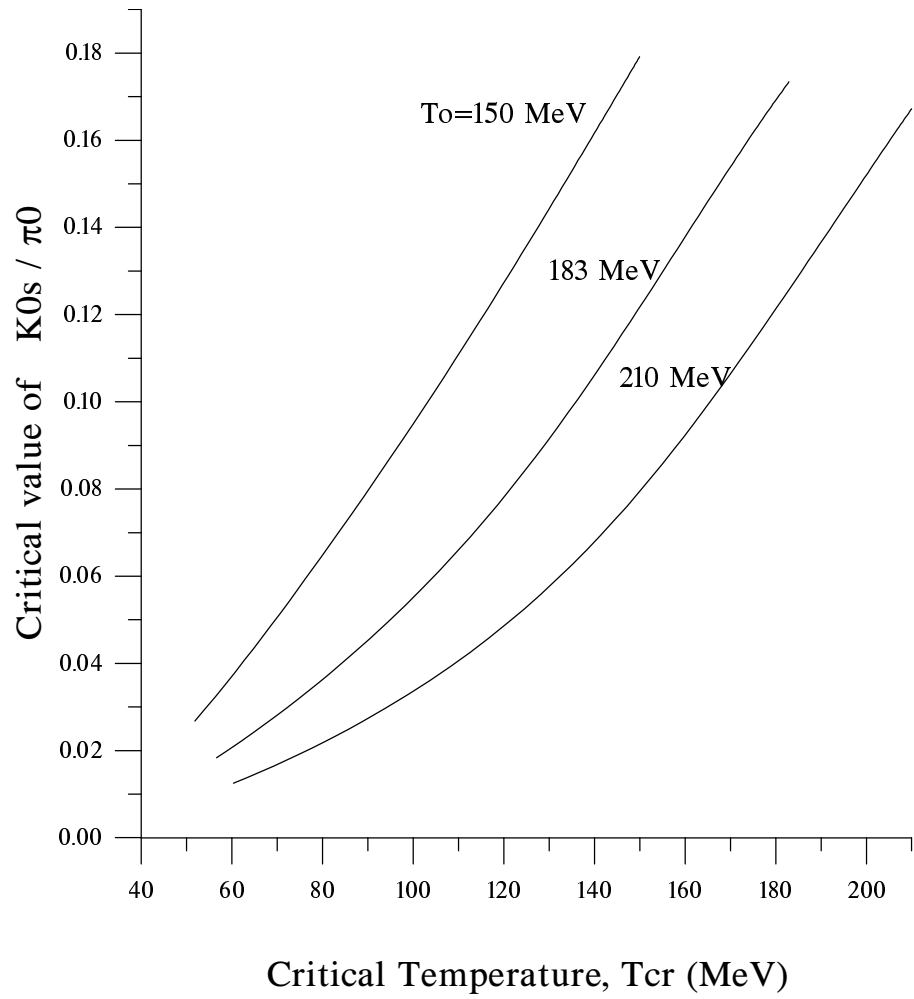
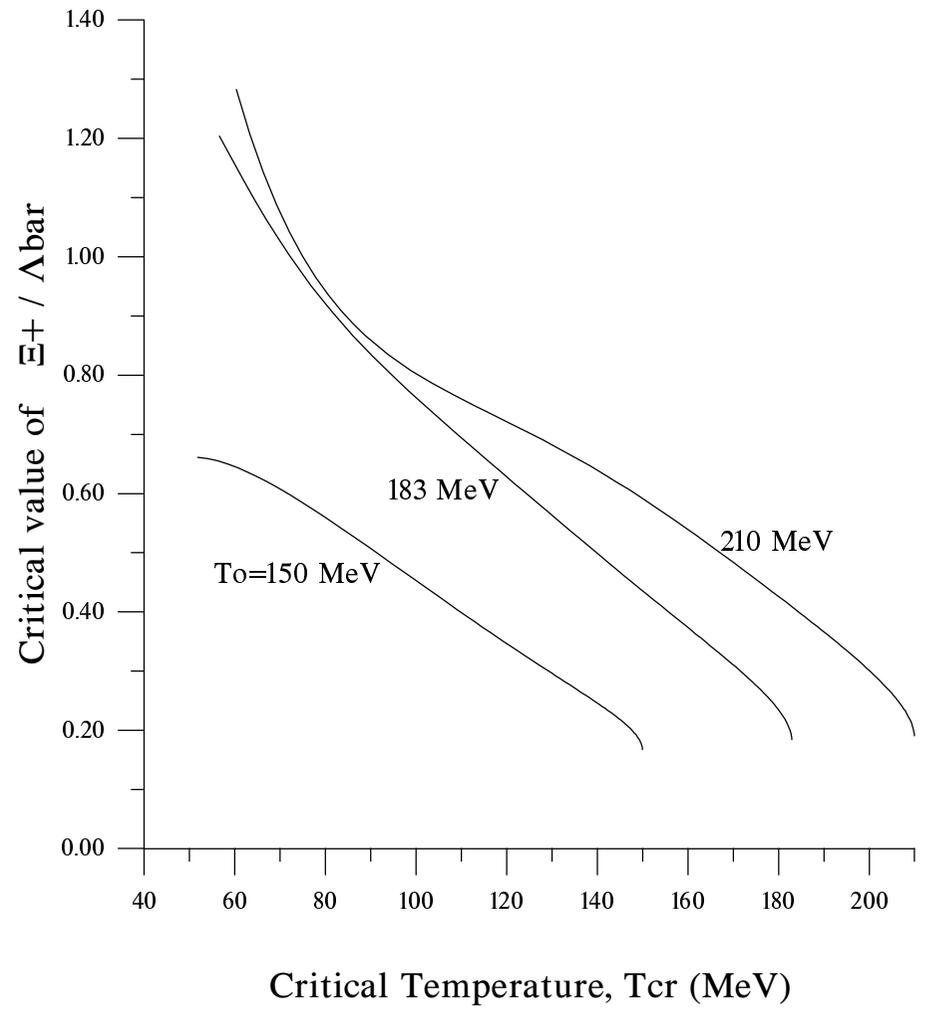
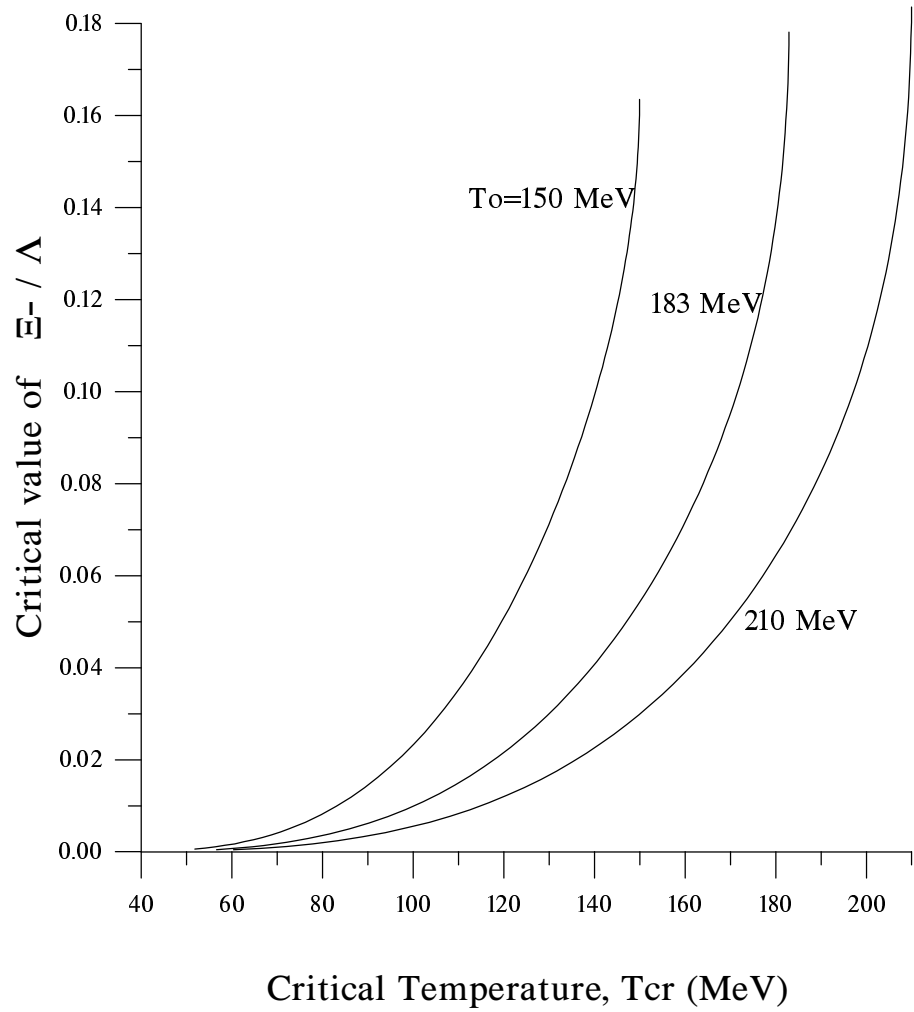
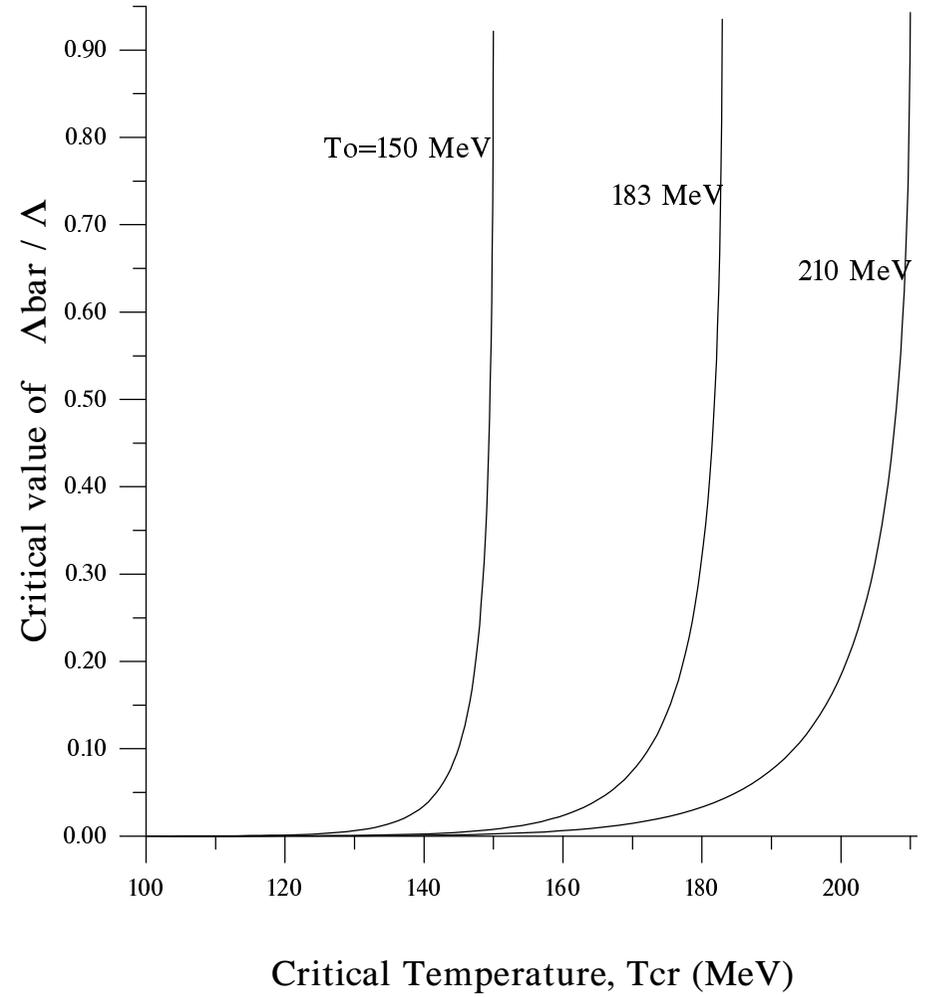
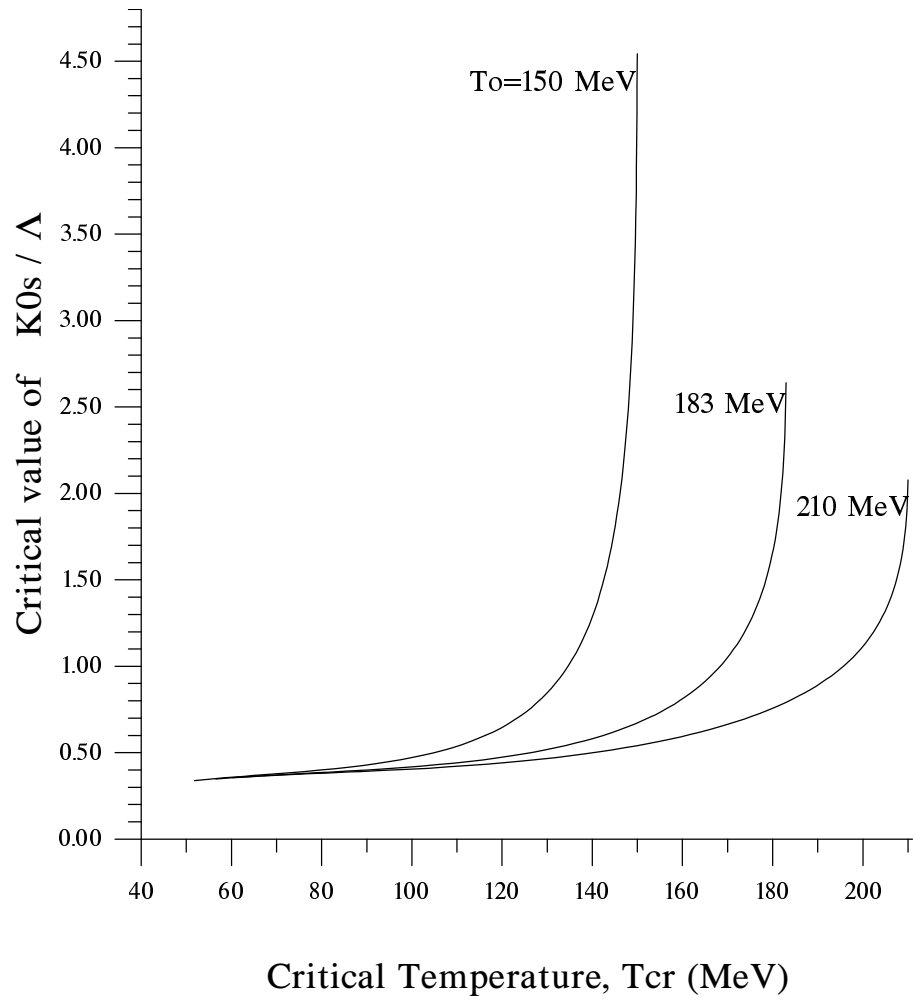
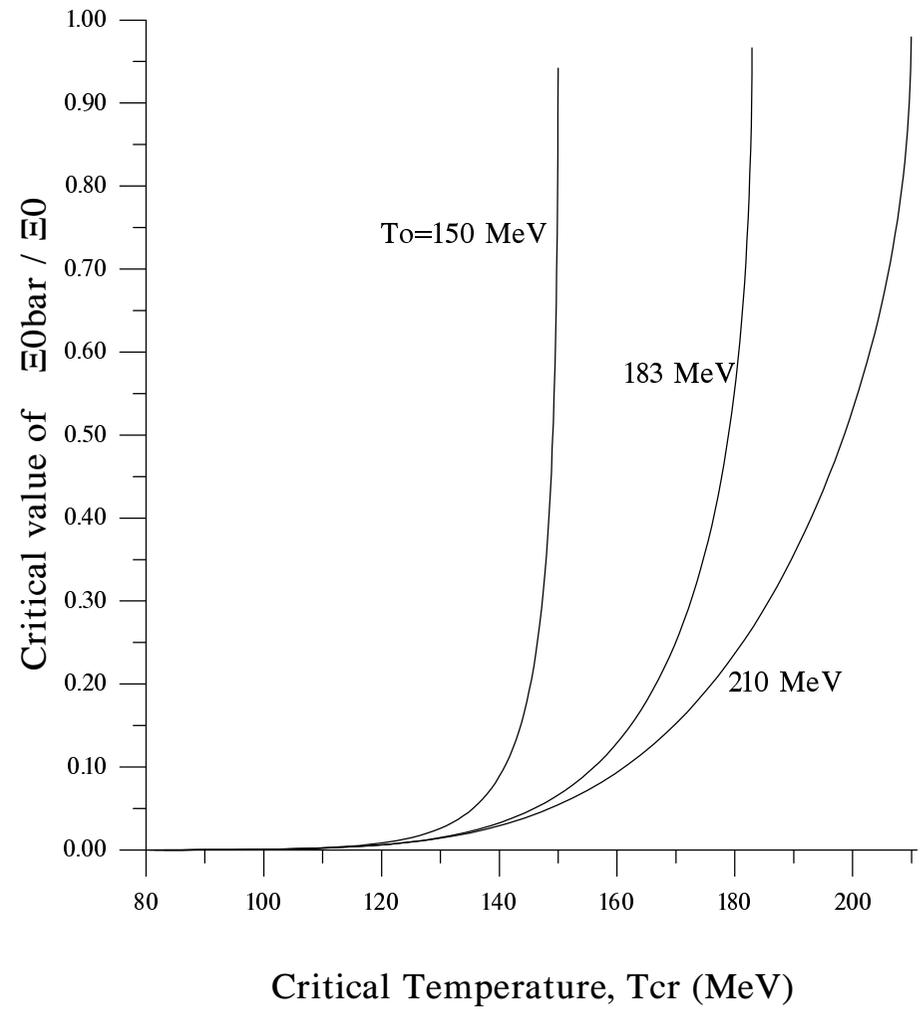
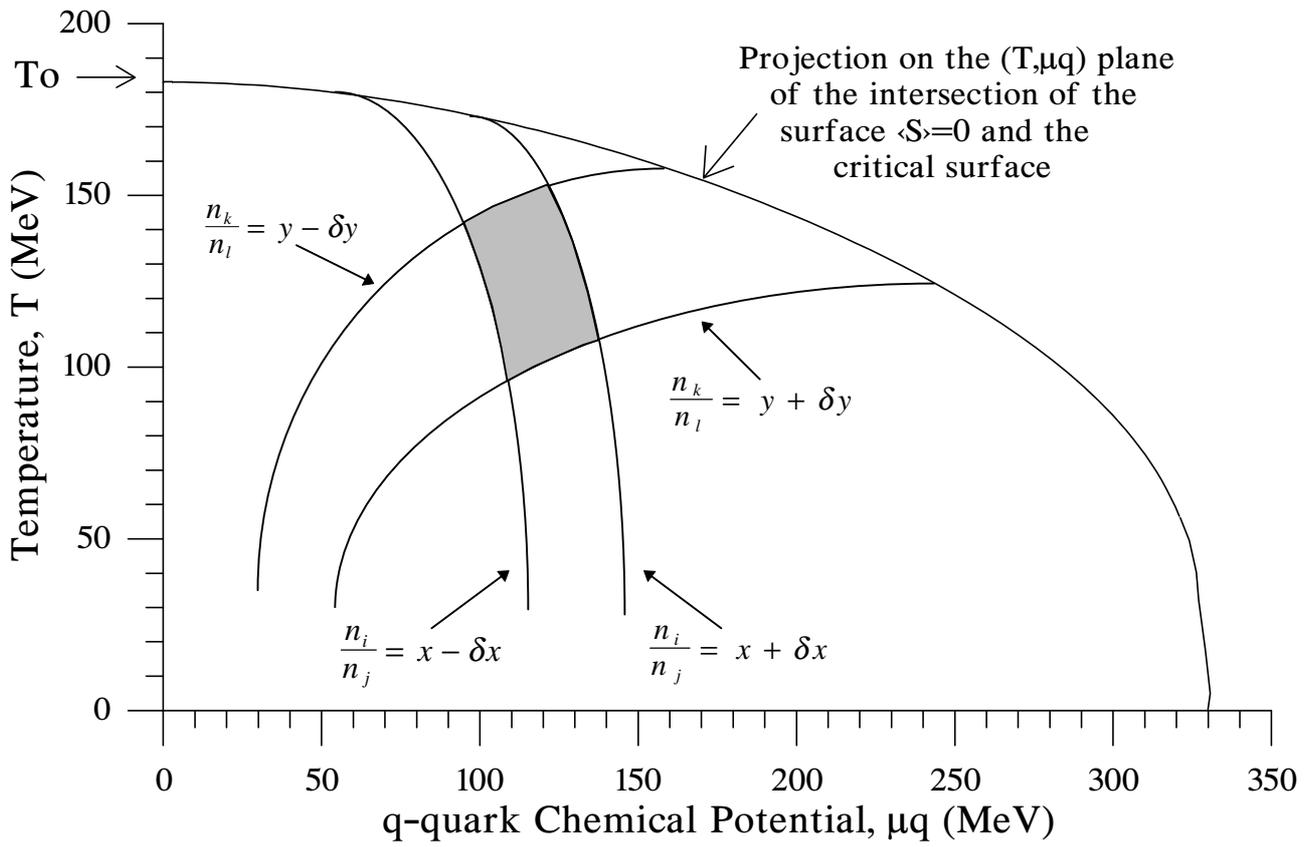
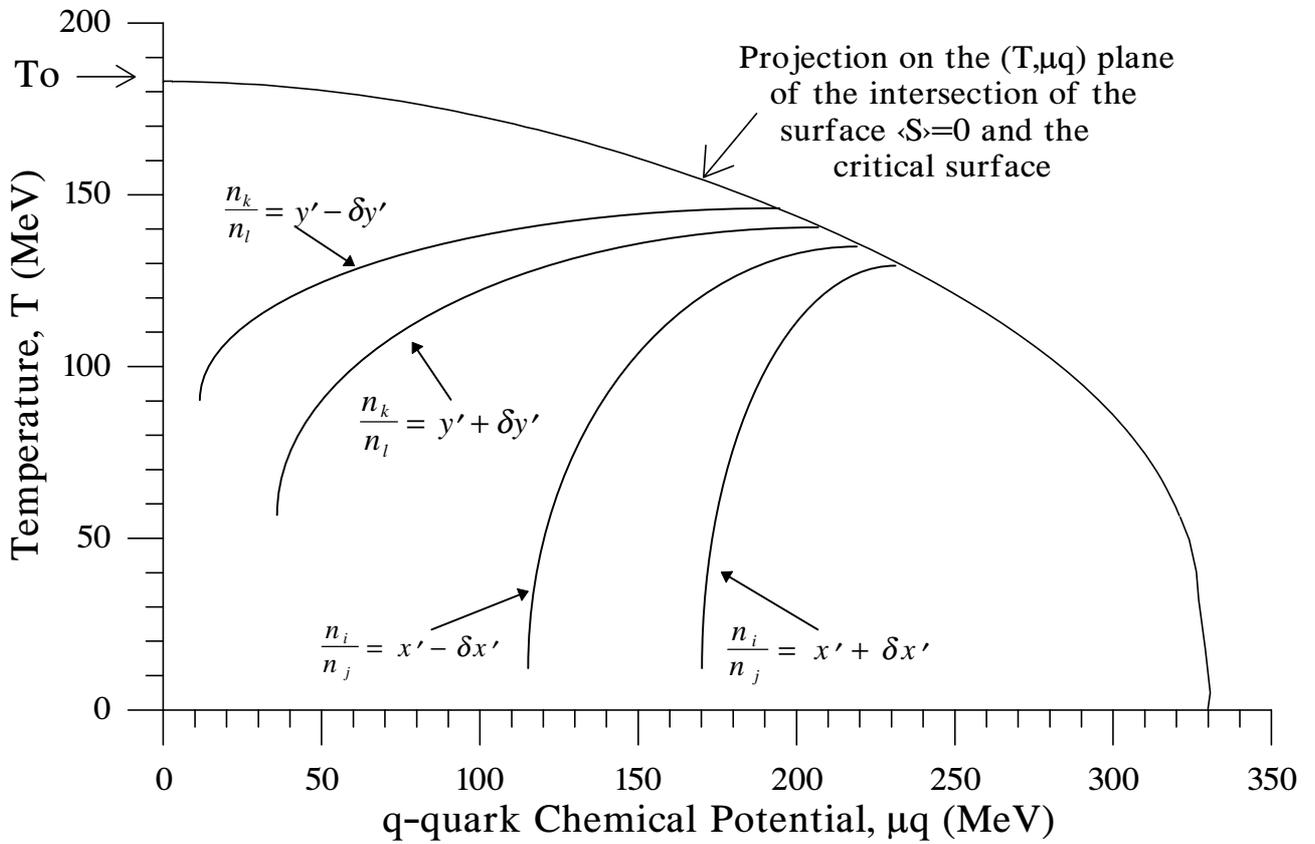
\end{document}